\documentclass[onecolumn]{IEEEtran}
\usepackage{cite}
\usepackage{amsmath,amssymb,amsfonts}
\usepackage{algorithmic}
\usepackage{graphicx}
\usepackage{epstopdf}
\usepackage{textcomp}
\usepackage{multirow}
\usepackage{booktabs}
\usepackage{amsmath}
\def\BibTeX{{\rm B\kern-.05em{\sc i\kern-.025em b}\kern-.08em
    T\kern-.1667em\lower.7ex\hbox{E}\kern-.125emX}}

\begin{document}
\title{Investigation of Numerical Dispersion with Time Step of The FDTD Methods: Avoiding Erroneous Conclusions}
\author{Yu Cheng, Guangzhi Chen, Xiang-Hua Wang, and Shunchuan Yang
\thanks{This paper is a preprint of a paper submitted to IET Microwaves, Antennas Propagation. If
	accepted, the copy of record will be available at the IET Digital Library.}
\thanks{Y. Cheng, G. Chen and S. Yang are with the School of Electronic and Information Engineering, Beihang University, Beijing, 100083, China (e-mail: yucheng@buaa.edu.cn, dazhihaha@buaa.edu.cn, scyang@buaa.edu.cn.)}
\thanks{X. H. Wang is with the School of Science, Tianjin University of Technology and Education, Tianjin, 300222, China (e-mail:  xhwang199@outlook.com.)}
}

\maketitle

\begin{abstract}
It is widely thought that small time steps lead to small numerical errors in the finite-difference time-domain (FDTD) simulations. In this paper, we investigated  how time steps impact on numerical dispersion of two FDTD methods including the FDTD(2,2) method and the FDTD(2,4) method. Through rigorously analytical and numerical analysis, it is found that small time steps of the FDTD methods do not always have small numerical errors. Our findings reveal that these two FDTD methods present different behaviours with respect to time steps: (1) for the FDTD(2,2) method, smaller time steps limited by the Courant-Friedrichs-Lewy (CFL) condition increase numerical dispersion and lead to larger simulation errors; (2) for the FDTD(2,4) method, as time step increases, numerical dispersion errors first decrease and then increase. Our findings are also comprehensively validated from one- to three-dimensional cases through several numerical examples including wave propagation, resonant frequencies of cavities and a practical engineering problem.  
\end{abstract}


\section{Introduction}
\label{sec:introduction}
\IEEEPARstart{T}{he} finite-difference time-domain (FDTD) method is one of the most widely used numerical methods to solve the practical electromagnetic problems, like scattering from electrically large and multiscale objects \cite{SCATTERING, LARGE,MULTISCALE}, integrated circuits \cite{FDTDCIRCUITS, FDTDCIRCUITS2,FDTDCIRCUITS3,FDTDCIRCUITS4}, electromagnetic compatibility (EMC) \cite{FDTDEMC,FDTDEMC2,FDTDEMC3}, electromagnetic interference (EMI) \cite{FDTDEMI,FDTDEMI2,FDTDEMI3}, due to its easy implementation, robustness,  strong capability of handling complex media and highly efficient parallel computation \cite{PARALLELFDTD1,PARALLELFDTD2}.  

However, the accuracy of the FDTD methods can be affected by several factors, such as numerical dispersion, mesh size, staircase errors, time steps, and so on. Numerical dispersion is one of the main factors that must be taken into consideration in FDTD methods, which implies that wavenumber of electromagnetic waves in the Yee's grid does not linearly depend on frequency. Numerical dispersion of the FDTD method and its variations, like high order FDTD methods \cite{HIGHORDER,HIGHORDERLIU,HIGHORDERTAN}, the alternatively-direction-implicit (ADI) FDTD methods \cite{ADI99,ADICHEN}, the locally 1-D (LOD) FDTD methods \cite{LOD}, are extensively investigated in a substantial literature  \cite{DISPADI,DISPLOD,DISFDTDLOSSY,DISLEAPFDTD,DISLOD2,DISFDTD243,DISFDTD}. Most of them focus on investigation of the relationship between numerical dispersion and propagation angles both in $\theta$ and $\phi$, mesh size, spatial distributions. Effects of time steps on numerical dispersion of the explicitly marching FDTD methods are seldom studied since one might think that it is relative simple and small time steps can reduce numerical dispersion. It is more common for the implicitly unconditionally stable FDTD methods to numerically discuss numerical dispersion with respect to time steps \cite{ADI99,ADICHEN,LOD,LEAPFROGADI}, since the Courant-Friedrichs-Lewy (CFL) condition is removed and large time steps are possible. It is found that for certain implicit FDTD methods, like the ADI-FDTD method and the LOD-FDTD method, numerical dispersion indeed decreases with smaller time steps \cite{ADI99,ADICHEN,LOD,LEAPFROGADI}.
  
 Then, several approaches to minimize numerical dispersion are proposed to obtain more accurate results without significantly increasing computational costs.  One simple but suboptimal option is to use fine enough grid in the FDTD simulations. It indeed increases the accuracy \cite{TAFLOVE}, however, inevitably with significantly increasing computational resources in terms of memory and CPU time since quite small mesh has to be used to capture fine geometric features, like wires and slots. Another type is to use optimized updating coefficients or artificial anisotropy in the time-marching formulations  \cite{FDTDOPTIMIZED,ADIOPTIMIZED,LODOPTIMIZED,FDTDOPTIMZIED2,FDTD24TIMESTEP}. In essence, through minimize numerical dispersion of different FDTD methods, the accuracy can be greatly improved without decreasing mesh sizes. Last but not least, high order finite-difference schemes are used to approximate the partial differential derivatives in the temporal and/or spatial domain \cite{HIGHORDER,HIGHORDERLIU,HIGHORDERTAN}, which show that significant accuracy improvement can be obtained.

It is easy to take for granted that a smaller time step leads to smaller numerical dispersion errors (NDEs) and then more accurate results in the FDTD simulations since 
\begin{equation}{\label{DEFPARI}}
\frac {\partial f} {\partial t} =  \lim\limits_{\Delta t \rightarrow 0} \frac{f^{n+1} - f^{n}}{\Delta t}.
\end{equation}
It is obvious that a larger $\Delta t$ would have larger approximation errors as shown in (\ref{DEFPARI}).  One may erroneously conclude that to obtain acceptable numerical results in the FDTD simulations, a small time step is much preferred. As shown in \cite{ADI99,ADICHEN,LOD,LEAPFROGADI}, although the implicit updating methods are unconditionally stable, time steps can not be arbitrarily large if high accurate results are required.  Otherwise, numerical dispersion can severely degenerate the accuracy and even totally unacceptable. It implies that long simulation time has been inevitable because of small time steps used in the FDTD methods. Therefore, large time step and small NDE seem to be contradictive. However, {\textit{is it really true}}? In this paper, we would report several findings upon effects of time steps in two classic FDTD methods upon numerical dispersion. Our findings reveal that smaller time steps do not always necessarily decrease numerical dispersion. On the contrary, it may lead to larger errors. Numerical dispersion of two typical FDTD methods including the FDTD(2,2) method and the FDTD(2,4) method are analytically and numerically studied in terms of various time steps. Then, we discussed optimal time steps for those methods with minimized numerical dispersion. 
 
 This paper is organized as follows. In Section II, the analytical numerical dispersion relationships of the two FDTD methods are briefly summarized. In Section III, effects of time steps of the two FDTD methods upon numerical dispersion are analytically  investigated and then NDEs are discussed. In Section IV, numerical studies upon numerical dispersion are carried out to validate our analytical analysis. In Section V, several numerical examples including wave propagation, resonant frequencies of cavities and a practical  engineering problem further validate our findings. At last, we draw some conclusions in Section VI.

\section{Numerical Dispersion Formulations for \\ Two FDTD Methods}
\subsection{Numerical Dispersion of the FDTD(2, 2) Method}
Without loss of generality, a homogenous, lossless, isotropic medium and uniform mesh is considered in our study. Therefore, the analytical numerical dispersion of the FDTD(2,2) method \cite{TAFLOVE} can be expressed as 
\begin{equation}\label{FDTDDISPERSIONFDTD22}
{\left[ {\frac{{\sin \left( {\omega \Delta t/2} \right)}}{{c\Delta t}}} \right]^2} = \sum\limits_{\xi  = x,y,z}^{} {{{\left[ {\frac{{\sin \left( {{{\tilde k}_\xi }\Delta \xi /2} \right)}}{{\Delta \xi }}} \right]}^2}} ,
\end{equation}
where $ {\tilde{k}_x} = \tilde{k}\sin \theta \cos \varphi $, $ \tilde{k}_y = \tilde{k}\sin \theta \sin \varphi $, $ {\tilde{k}_z} = \tilde{k}\cos \theta $  are numerical wavenumbers in the $x$, $y$ and $z$ direction, respectively, $\theta$ and $\phi$ are the azimuth and zenith angles, $\tilde{k}$ is the numerical wavenumber of electromagnetic waves in the Yee's grid, $\Delta x$, $\Delta y$ and $\Delta z$ are mesh sizes in the $x$, $y$ and $z$ direction, respectively. $\omega$ denotes the angular frequency and $\Delta t$ is the time step. 
 
To obtain stable numerical solutions for the explicit FDTD (2,2) method, time steps must satisfy the CFL condition \cite{TAFLOVE}, expressed as
\begin{equation}\label{CFL}
\Delta t \le \frac{{\sqrt {\varepsilon \mu } }}{{\sqrt {{{\left( {\Delta x} \right)}^{ - 2}} + {{\left( {\Delta y} \right)}^{ - 2}}{\rm{ + }}{{\left( {\Delta z} \right)}^{ - 2}}} }}.
\end{equation}

For convenience in our following derivation, $S_{fdtd(2,2)}$ is defined as
\begin{equation}\label{CFLS}
S_{fdtd(2,2)}=\frac{\Delta t}{\Delta t_{max\_{fdtd(2,2)}}},
\end{equation}
where $\Delta t_{max\_{fdtd(2,2)}}$ is the maximum time step defined by the CFL condition in (\ref{CFL}).

\subsection{Numerical Dispersion Relationship of the FDTD(2,4) Method}
The analytical numerical dispersion of the FDTD(2,4) method \cite{TAFLOVE} is written as
\begin{equation}\label{FDTDDISPERSIONFDTD24}
\begin{aligned}
{\left[ {\frac{{\sin (\frac{{\omega \Delta t}}{2})}}{{c\Delta t}}} \right]^2} = \sum\limits_{\xi  = x,y,z}^{} {{{\left[ {\frac{{27\sin \left( {\frac{{{k_\xi }\Delta \xi }}{2}} \right) - \sin \left( {\frac{{3{k_\xi }\Delta \xi }}{2}} \right)}}{{24\Delta \xi }}} \right]}^2}} 
\end{aligned},
\end{equation}

 Then, the CFL condition for the FDTD(2,4) method \cite{TAFLOVE} is expressed as 
\begin{equation}\label{CFL24}
\Delta t \le \frac{6}{7}\frac{{\sqrt {\varepsilon \mu } }}{{\sqrt {{{(\Delta x)}^{ - 2}} + {{(\Delta y)}^{ - 2}} + {{(\Delta z)}^{ - 2}}} }}.
\end{equation}

We further define $S_{fdtd(2,4)}$ as
\begin{equation}\label{CFLS24}
S_{fdtd(2,4)}=\frac{\Delta t}{\Delta t_{max\_{fdtd(2,4)}}},
\end{equation}
where $\Delta t_{max\_{fdtd(2,4)}}$ is the maximum time step constrained by the CFL condition in (\ref{CFL24}).

\section{Analytical Analysis of Numerical Dispersion of the FDTD Methods}

In this section, we theoritically discuss effects of time steps on numerical dispersion of the two FDTD methods. All the subscripts of $S_{fdtd(2,2)}$ and $S_{fdtd(2,4)}$ are suppressed for brevity.

\subsection{The FDTD(2,2) Method}

 {\bf{Remark}} 1: $\tilde k \ge k$, $\Delta t \in  (0, t_{max\_fdtd(2,2)}]$. 
  
 Since the explicit expression of $\tilde k$ can not be directly obtained from  (\ref{FDTDDISPERSIONFDTD22}), the parameter scanning method {\cite{PARASCANNTING,PARASCANNTING2}} is used to study the relationship between $\tilde k$ and $k$. We scanned $\theta \in [0, \pi]$, $\phi \in [0, 2\pi]$ with other possible parameters in (\ref{FDTDDISPERSIONFDTD22}). This above statement always holds true. Although it is not mathematically rigorous, it should be enough for practical applications and our analysis.  Here, we plot the maximum $\tilde{k}$ with $f=5$ GHz, $\Delta x$ =  $\Delta y$ = $\Delta z$ = $6 \times {10^{ - 3}}$ m. Both $\theta \in [0, \pi]$, $\phi \in [0, 2\pi]$ are scanned and the maximum $\tilde{k}$ is obtained as shown in Fig. 1. It can be easy to find that when $\Delta t$ satisfies the CFL condition, namely $S\in(0, 1]$, $\tilde{k}$ is always larger than $k$.

\begin{figure}\label{first}	
	\centerline{\includegraphics[width=3.5in]{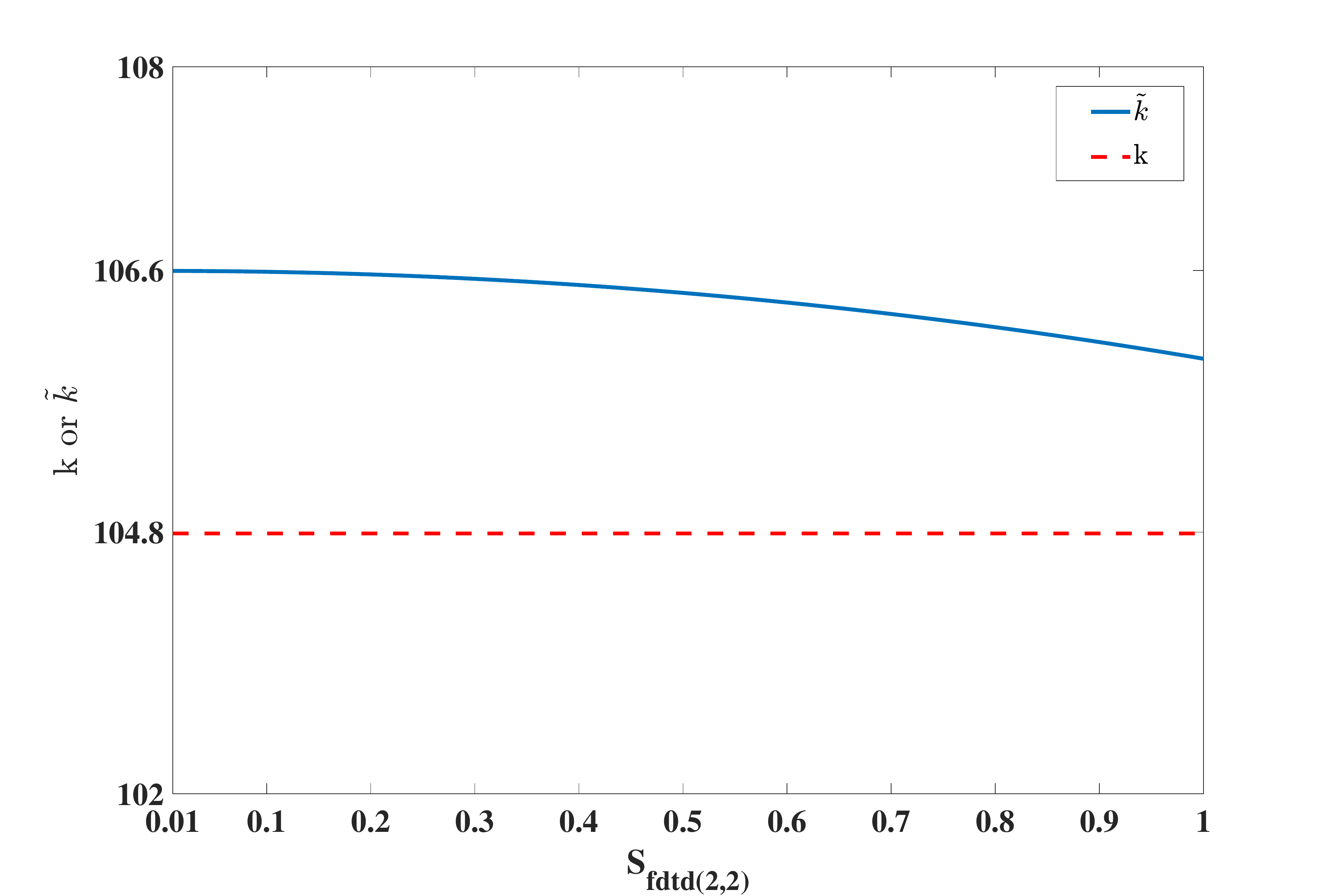}}
	
	\caption{The maximum $\tilde k$ and $k$ with $f=5$ GHz, $\Delta x$ =  $\Delta y$ = $\Delta z$ = $6 \times {10^{ - 3}}$ m for different $S_{fdtd(2,2)}$.
	}
\end{figure}
 
  {\bf{Lemma}} 1: For the FDTD(2,2) method, 
 \begin{equation}\label{FDTDDISPERSION}
 \frac{{d\tilde k}}{{dS}} \le 0,  \qquad S \in  (0, 1].
 \end{equation}
 
 {\bf{Proof:}} Since the explicit expression of $\tilde{k}$ is not available,  the implicit differentiation \cite{Calculus} is resorted to investigating the derivative of $\tilde{k}$ with respect to $S$ defined in (\ref{CFLS}). For convenience of derivations, the following symbols are defined as
\begin{equation}
	\text{LHS} = {\left[ {\frac{{\sqrt {\varepsilon \mu } \sin \left( {\omega \Delta t/2} \right)}}{{\Delta t}}} \right]^2},
\end{equation}
and
\begin{equation}
\text{RHS} = \sum\limits_{\xi  = x,y,z}{{{\left[ {\frac{{\sin \left( {{{\tilde k}_\xi }\Delta \xi /2} \right)}}{{\Delta \xi }}} \right]}^2}}.
\end{equation}
 Therefore, (\ref{FDTDDISPERSIONFDTD22}) is separated into LHS and RHS two parts. We first consider LHS. By assuming uniform mesh used in the simulations, $\Delta x = \Delta y = \Delta z = \Delta$, and substituting $
\omega  = {{2\pi }}/{{\lambda \sqrt {\varepsilon \mu } }}$ and (\ref{CFLS})
 into LHS,  we obtain
\begin{equation}\label{FDTDDISPERSIONm}
\text{LHS} = {\left[ {\frac{{\sqrt 3 }}{{S\Delta}}\sin \left( {\frac{{\pi S}}{{\sqrt 3 N}}} \right)} \right]^2},
\end{equation}
where $N$ is the count of sampling cells per wavelength, defined as $N = \lambda /\Delta$. For practical simulations, $N \ge 10$ should be satisfied to obtain acceptable accurate results \cite{TAFLOVE}.

With some mathematical manipulations, the derivative of LHS with respect to $S$ is obtained as 
\begin{equation}\label{LHSD}
\frac{{d({\rm{LHS}})}}{{dS}} = 6\sin \left( {\frac{{\pi S}}{{\sqrt 3 N}}} \right)\left[ {\frac{Q}{{\sqrt 3 N{S^3}{{(\Delta x)}^2}}}} \right],
\end{equation}  
where $Q = \left[ {\pi S\cos \left( {{{\pi S}}/{{\sqrt 3 N}}} \right) - \sqrt 3 N\sin \left( {{{\pi S}}/{{\sqrt 3 N}}} \right)} \right]$.
Since ${{\pi S}}/({{\sqrt 3 N}}) \in (0,{\pi }/{2})$, 
\begin{equation}{\label{SINN}}
\sin \left( \frac{\pi S} { {\sqrt 3 N} } \right)  > 0.
\end{equation}
Then, let's check the sign of $Q$ in (\ref{LHSD}). By taking its derivative, we get
\begin{equation}\label{LHSDD}
\frac{{dQ}}{{dS}} =  - \frac{{{\pi ^2}S}}{{\sqrt 3 N}}\sin \left( {\frac{{\pi S}}{{\sqrt 3 N}}} \right).
\end{equation}
Since $Q$ is a continous function, ${ - {\pi ^2}S\sin \left( {\pi S/\sqrt 3 N} \right)/\left( {\sqrt 3 N} \right) < 0}$, and ${\left. {\left[ {\pi \cos \left( {{{\pi S}}/{{\sqrt 3 N}}} \right)S - \sqrt 3 N\sin \left( {{{\pi S}}/{{\sqrt 3 N}}} \right)} \right]} \right|_{S = 0}} = 0$, we get

\begin{equation}\label{LHS2}
Q < 0.
\end{equation}
By considering (\ref{SINN}), (\ref{LHS2}) and other variable signs in (\ref{LHSD}), we can easily get 
\begin{equation}\label{LHSI}
\frac{{d\left( {{\rm{LHS}}} \right)}}{{dS}} < 0.
\end{equation}

 With a similar manner, the derivative of RHS can be expressed as
\begin{equation}\label{FDTDDISPERSIONRHS2}
\frac{{d({\rm{RHS}})}}{{dS}}{\rm{ = }}\frac{1}{{\Delta x}}\frac{{d\tilde k}}{{dS}}P,
\end{equation}
where $\tilde k = \omega /{\tilde v_p}$, where $\tilde v_p$ is the numerical phase velocity in the Yee's grid, and
\begin{equation}\label{PP}
P = \left[ \begin{array}{l}
\underbrace{\sin \theta \cos \varphi \sin \left( M \right) \cos \left( M \right)}_{\text{term1}}  {{ + }}\\
\underbrace{\sin \theta \sin \varphi \sin \left( M \right) \cos \left( M \right)}_{\text{term2}}{{ + }}\\
\underbrace{\cos \theta \sin \left( T \right)\cos \left( T \right)}_{\text{term3}}
\end{array} \right],
\end{equation}
with $M = {{{\pi a\sin \theta \cos \varphi }}/{N}}$, $T={{{\pi a\cos \theta }}/{N}}$, and $a = {c_0}/\tilde{v}_p$.

\begin{table}
	\centering
	\caption{Signs of (a) term1, (b) term2 and (c) term3 in (\ref{PP}).}
	\begin{minipage}[h]{0.48\linewidth}\label{FIG2A}
		\centerline{(a)}
		\centerline{\includegraphics[width=3in]{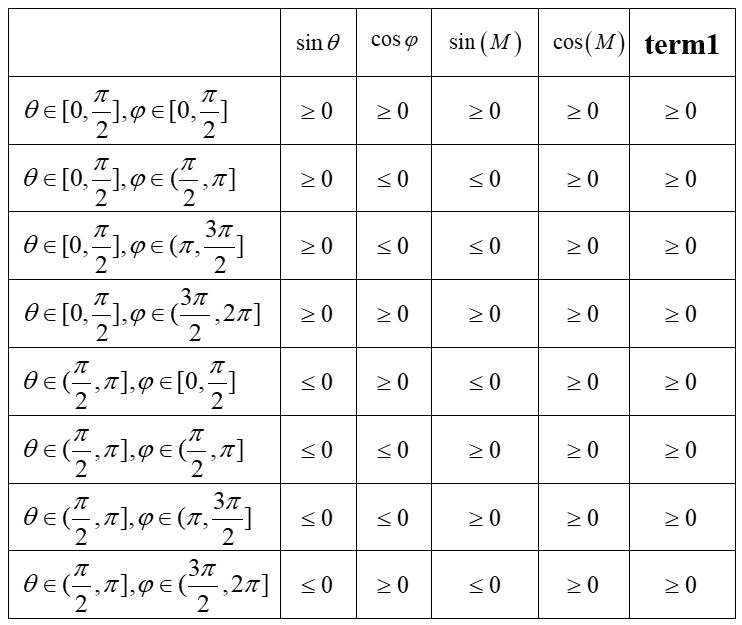}}			
	\end{minipage}

	\bigskip
	\centering
	\begin{minipage}[h]{0.48\linewidth}\label{FIG2B}
		\centerline{(b)}
		\centerline{\includegraphics[width=3in]{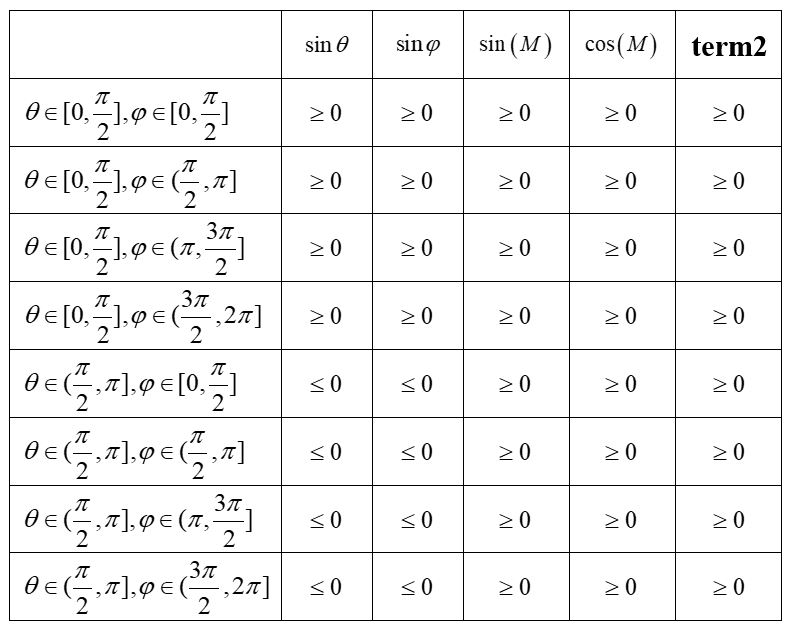}}		
	\end{minipage}
	
	\bigskip
	\centering
	\begin{minipage}[h]{0.48\linewidth}\label{FIG2C}
		\centerline{(c)}
		\centerline{\includegraphics[width=3in]{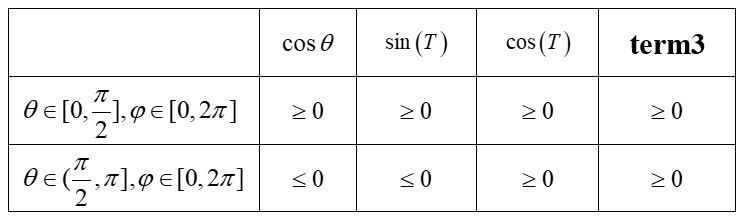}}	
	\end{minipage}
	
	\label{fig_2}
\end{table}

To further investigate signs of (\ref{PP}),  we divide its domain into several parts based on signs of their function values of each term. Note that the function domain is $\varphi \in [0, 2\pi]$ and $\theta \in [0, \pi]$. Therefore, we divide them into several parts to check their signs. The detailed results are summarized in Table I. Table I (a)-(c) correspond to three terms in (\ref{PP}).  To better illustrate it, we take term1 in (\ref{PP}) as an example. When $\theta  \in \left[ {0,\pi /2} \right]$ and $\varphi  \in \left[ {0,\pi /2} \right]$, it is evidently that $\sin \theta  \in \left[ {0,1} \right],\cos \varphi  \in \left[ {0,1} \right]$. Note that $(\pi a/N) \in \left( {0,\pi /2} \right)$, then $M \in [ {0,\pi /2} ]$. Therefore, we have $\sin \left( M \right) \ge 0$ and $\cos \left( M \right) \ge 0$, which implies that term1 is positive when $\theta  \in \left[ {0,\pi /2} \right]$ and $\varphi  \in \left[ {0,\pi /2} \right]$. The remaining situations can be obtained with similar manners. To sum them up, we can obtain the following inequality
\begin{equation}\label{inequ}
P \ge 0.
\end{equation}

By considering (\ref{LHSI}) and (\ref{inequ}), we get

\begin{equation}\label{fin}
\begin{footnotesize}
\frac{{d\tilde k}}{{dS}} = \frac{{d({\rm{LHS}})}}{{dS}} \frac{\Delta x}{P} \le 0.
\end{footnotesize}
\end{equation}

Evidently, {\bf{Lemma}} 1 is analytically proved and we can further analyze effect of time steps on the NDE of the FDTD(2,2) method.

From {\bf{Remark}} 1 and {\bf{Lemma}} 1,  we can easily find that the numerical wavenumber $\tilde{k}$ monotonically decreases as $S$ in the FDTD(2,2) method increases. When time steps grow larger and are bounded by (\ref{CFL}), the numerical phase velocity $\tilde{v}_p$ of electromagnetic waves is closer to its physical counterpart $c_0$, and then NDE caused by the numerical dispersion would be relatively smaller. Therefore, the NDE reaches its minimum value when $S$ gets its maximum value defined by (\ref{CFL}). 

With similar procedures, we can investigate effects of time steps upon numerical dispersion for the FDTD(2,4) method.

\subsection{The FDTD(2,4) Method}

{\bf{Remark}} 2: $\exists t\_  \in (0, t_{max\_fdtd(2,4)}]$, we have $\tilde k \ge k$, $\Delta t \in  (0, t\_ ], \tilde k \le k$, $\Delta t \in  (t\_ , t_{max\_fdtd(2,4)}]$.

{\bf{Lemma}} 2: For the FDTD(2,4) method, 
\begin{equation}\label{FDTD24DISPERSION}
\frac{{d\tilde k}}{{dS}} \le 0,  \qquad S \in  (0, 1].
\end{equation}

From {\bf{Remark}} 2 and {\bf{Lemma}} 2,  the NDE $\tilde{k}$ firstly decreases as the $S$ of the FDTD(2,4) method get larger. When  $\Delta t \in  (t\_ , t_{max\_fdtd(2,4)}]$, the NDE becomes larger. Therefore, the NDE reaches its minimum value at $t = t\_$.

\section{Numerical Analysis of Numerical Dispersion of the FDTD Methods}
\subsection{The FDTD(2, 2) Method}
To comprehensively investigate the numerical dispersion of the FDTD (2,2) method, the numerical phase velocity $\tilde{v}_p$ is calculated from (\ref{FDTDDISPERSIONFDTD22}).  In our numerical studies, the frequency is 5 GHz and $\Delta x$ =  $\Delta y$ = $\Delta z$ = $6 \times {10^{ - 3}}$ m, which corresponds to $\lambda/10$.  In all other simulations, the same parameters are used, otherwise stated.  

 As shown in Fig. 2,  $\tilde{v}_p$ changes periodically over  $\varphi $ for a fixed time step $S_{fdtd(2,2)}$ when $\theta = 90^o$. However,  $\tilde{v}_p$ increases as $S_{fdtd(2,2)}$ becomes larger with fixed $\phi$ values, which implies that $\tilde{v}_p$ approaches $c_0$ as time steps grow larger. Therefore, the NDE would become smaller as time steps get larger, which agrees with our analytical analysis in the previous section.  Fig. 3 shows  $\tilde{v}_p$ with respect to $S_{fdtd(2,2)}$ when $\phi = 90^o$. Similar statements can be also obtained. 
 
Fig. 4 illustrates $\tilde{v}_p$ with different $S_{fdtd(2,2)}$ with respect to $\varphi $ and $\theta$. It is easy to find that each $\tilde{v}_p$ surface does not intersect each other. Therefore, we can conclude that $\tilde{v}_p$ approaches $c_0$ as $S_{fdtd(2,2)}$ becomes larger.  The numerical dispersion results show that the optimum time step of the FDTD(2,2) method is the largest time step defined by the CFL condition, which also agrees with our theoretical analysis.

\begin{figure}\label{FDTDANA}	
	\centerline{\includegraphics[width=3.5in]{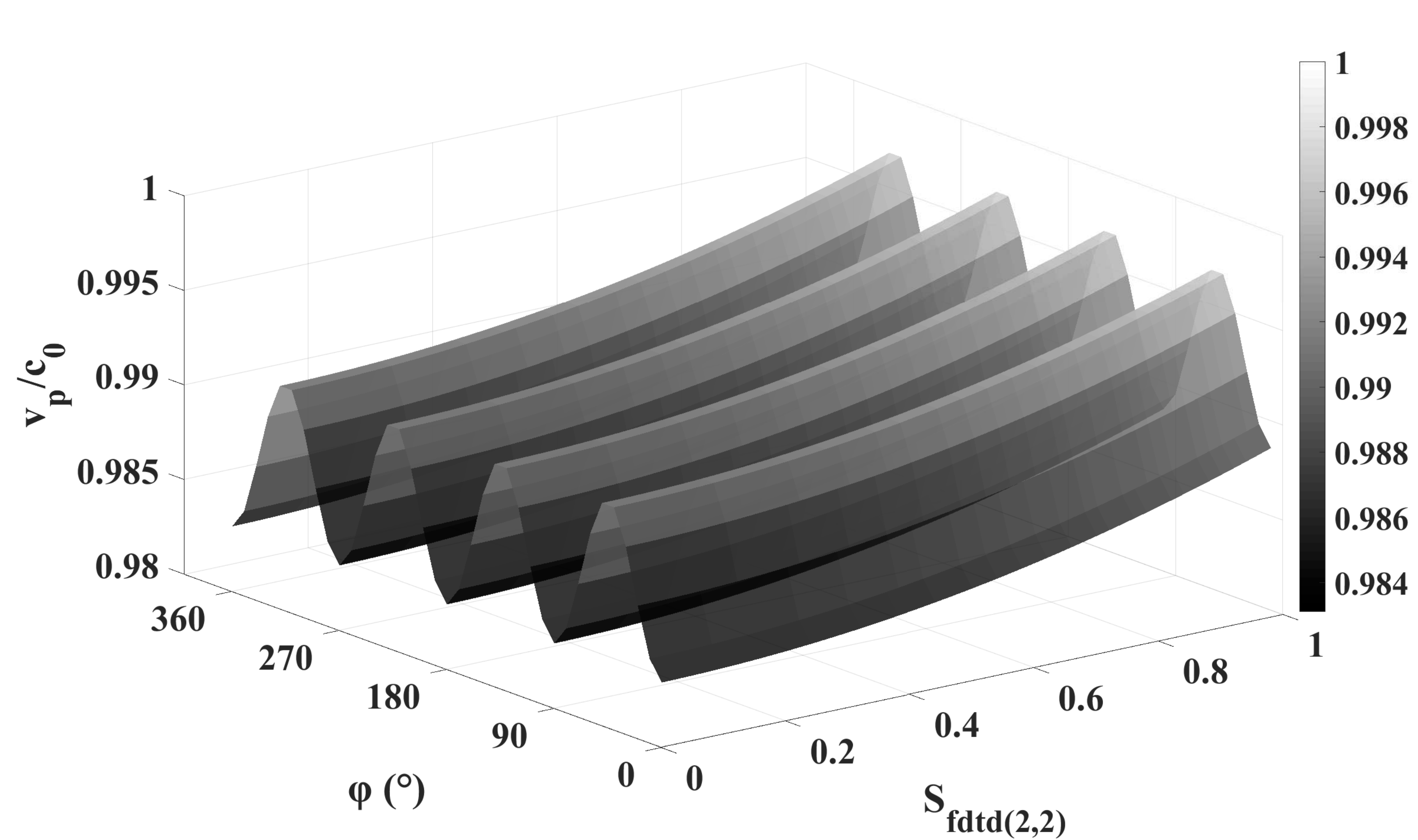}}
	\caption{$\tilde{v}_p/ c_0$  of the FDTD(2,2) method with respect to $S_{fdtd(2,2)}$ and $\varphi $  when $\theta  = {90^ \circ }$, $f = 5$ GHz and $\lambda /\Delta = 10$.}
\end{figure}

\begin{figure}\label{FDTDTHE}	
	\centerline{\includegraphics[width=3.5in]{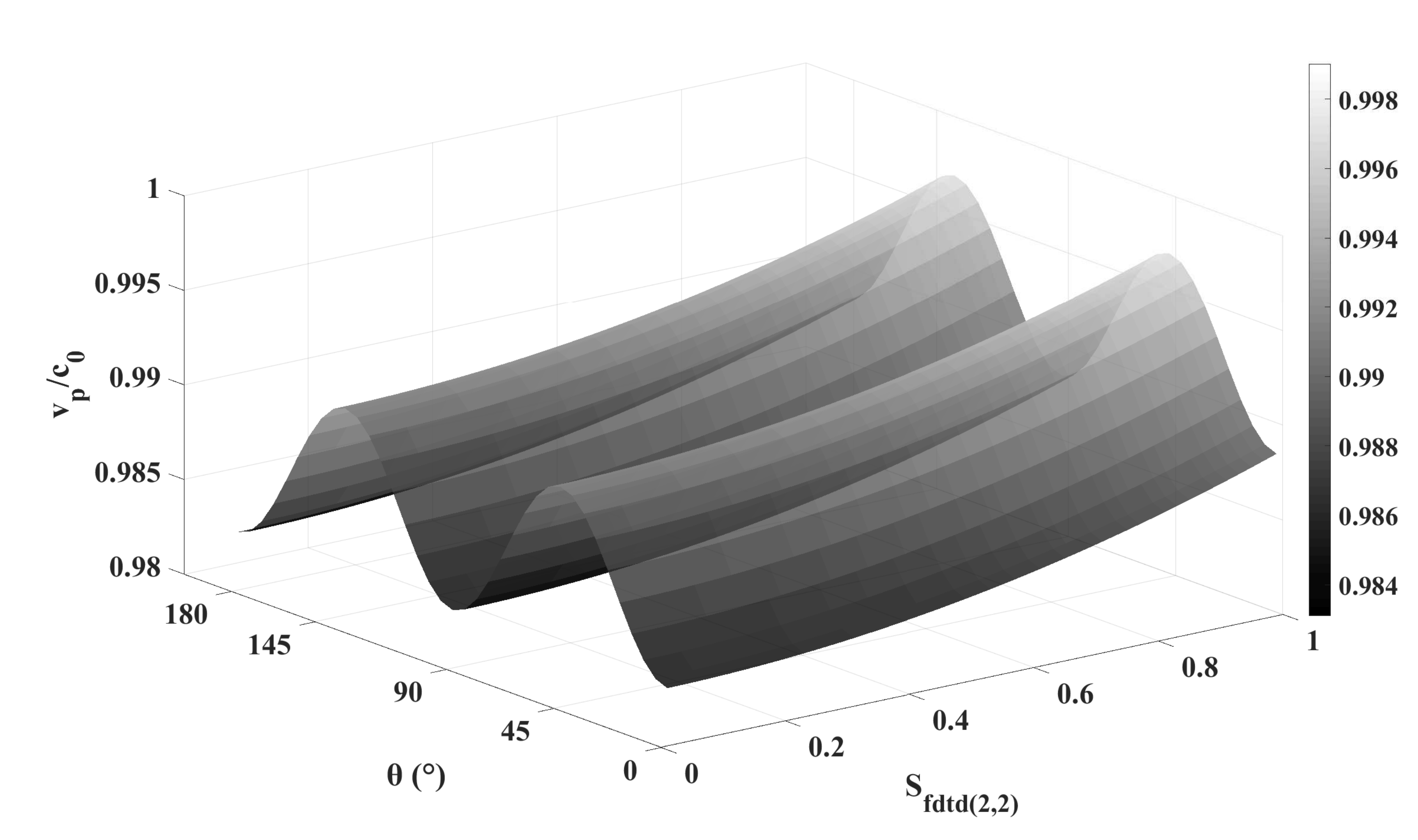}}
	\caption{$\tilde{v}_p/ c_0$  of the FDTD(2,2) method with respect to $S_{fdtd(2,2)}$ and $\theta $  when $\varphi  = {90^ \circ }$, $f = 5$ GHz and $\lambda /\Delta = 10$.}
\end{figure}

\begin{figure}\label{fdtd_}	
	\centerline{\includegraphics[width=3.5in]{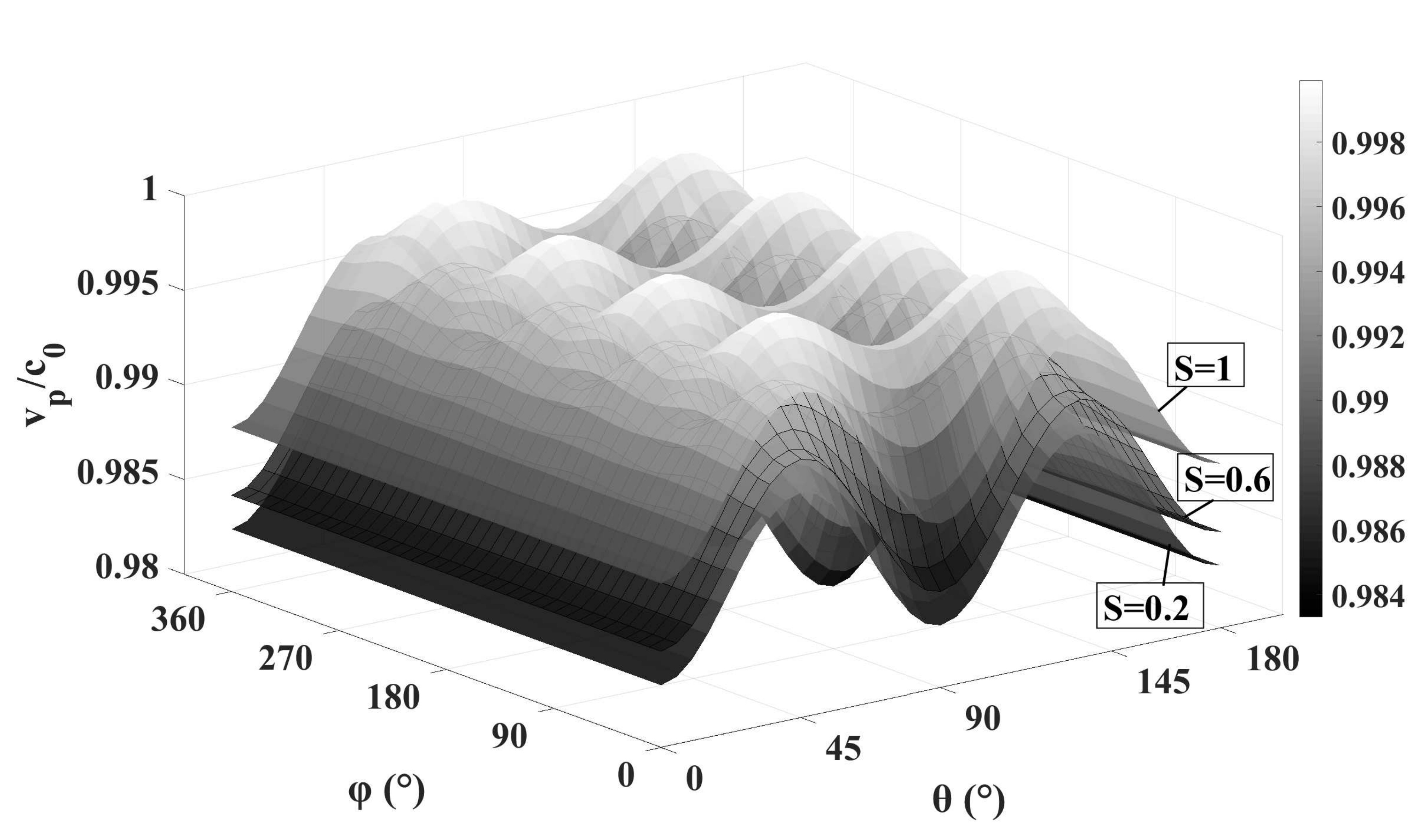}}
	\caption{$\tilde{v}_p/ c_0$  of the FDTD(2,2) method with respect to $S_{fdtd(2,2)}$, $\theta $ and $\varphi$ when $f = 5$ GHz and $\lambda /\Delta  = 10$. The top surface corresponds to $S_{fdtd(2,2)} = 1$ .}
\end{figure}

\subsection{The FDTD(2,4) Method}
$\tilde{v}_p/c_0$ of the FDTD (2,4) method is also calculated to investigate its numerical dispersion with different $S_{fdtd(2,4)}$. As shown in Fig. 5, $\tilde{v}_p/c_0$ changes periodically with respect to $\varphi$ when $\theta = 90^\circ$.  $\tilde{v}_p$ approaches to $c_0$ at the beginning when $S_{fdtd(2,4)}$ becomes large, and then becomes larger than $c_0$ as $S_{fdtd(2,4)}$ further increases. Therefore, the NDE of the FDTD (2,4) method firstly decreases and then increases over $S_{fdtd(2,4)}$. The statements obtained in Fig. 6 are similar to those of Fig. 5. They agree with our theoretical analysis.

Here, we provide a numerical method to calculate the optimum time step of the FDTD(2,4) method. The optimum time step is defined as   
\begin{equation}\label{FDTDtwofour}
\Delta t \rightarrow \text{min}{\left\{ {\int_0^{2\pi } {\int_0^\pi  {\left| {\tilde k - {k}} \right|} d\theta d\varphi } } \right\}},
\end{equation}
which can make difference between the numerical wavenumber and its analytical counterpart with $\theta \in [0, \pi]$ and $\varphi \in [0, 2\pi]$ minimum.

\begin{figure}\label{twofour}	
	\centerline{\includegraphics[width=3.5in]{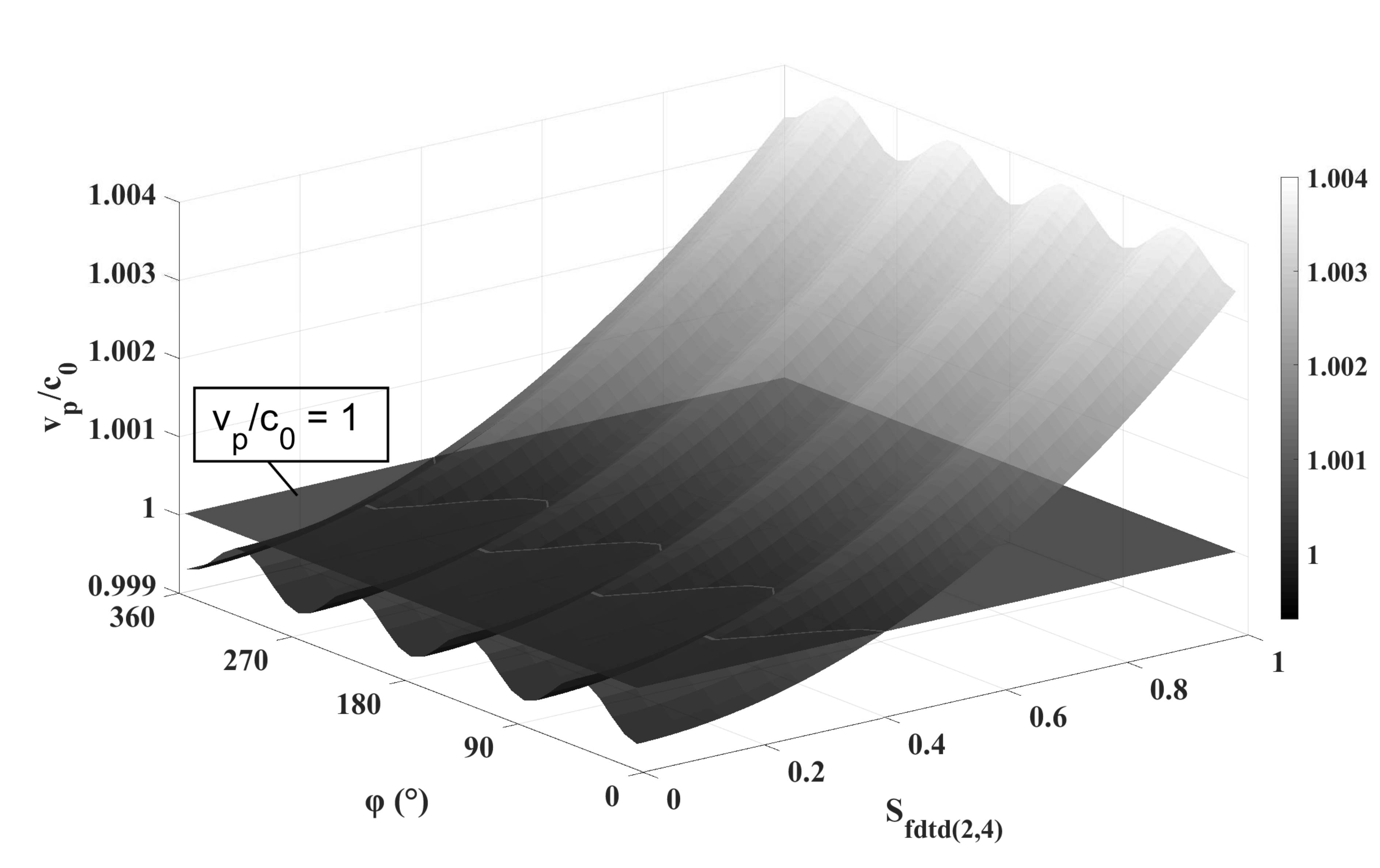}}
	\caption{$\tilde{v}_p/ c_0$  of the FDTD(2,4) method with respect to $S_{fdtd(2,4)}$ and $\varphi $  when $\theta  = {90^ \circ }$, $f = 5$ GHz and $\lambda /\Delta = 10$.}
\end{figure}

\begin{figure}\label{twofourtheta}	
	\centerline{\includegraphics[width=3.5in]{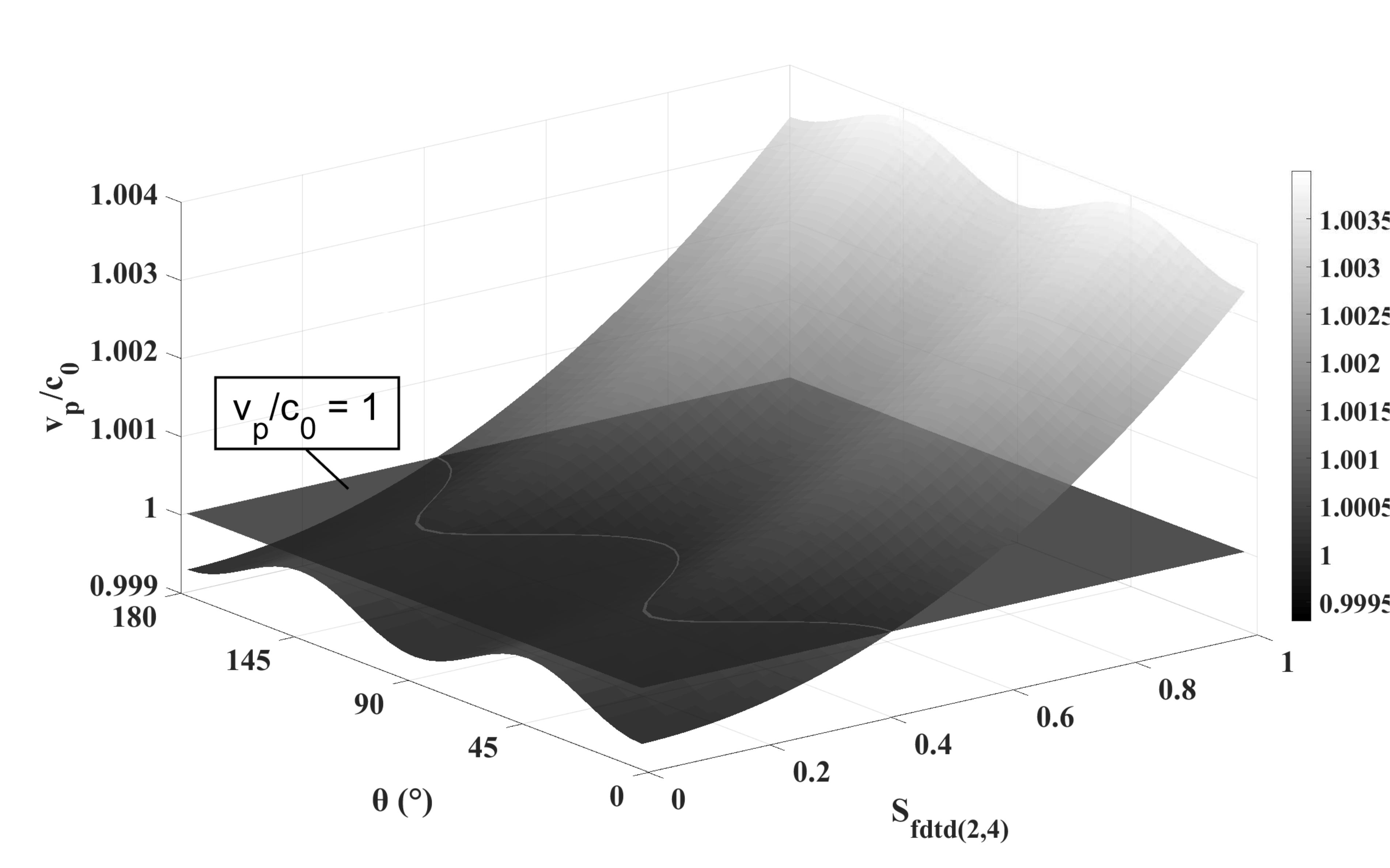}}
	\caption{$\tilde{v}_p/ c_0$  of the FDTD(2,4) method with respect to $S_{fdtd(2,4)}$ and $\theta $  when $\varphi  = {90^ \circ }$, $f = 5$ GHz and $\lambda /\Delta = 10$.}
\end{figure}

\section{Numerical Results and Discussion}
In this section, several numerical examples  including wave propagation, resonant frequencies of cavities and a practical engineering problem are carried out to validate our previous analysis. The relative error is defined as
\begin{equation}
\text{RE} = \left| \frac{f^{\text{ref}} -  f^{\text{cal}}}{f^{\text{ref}}}  \right |,
\end{equation}
where $f^{\text{ref}}$ and $f^{\text{cal}}$ denotes the reference and calculated values, respectively.

\subsection{The One Dimensional Case}
A wave propagation example is used to verify the accuracy of the two FDTD methods with different time steps in one dimensional case.  A Gaussian function with the waveform $\phi \left( t \right) = {e^{ - 16{{\left( {0.7 - ct} \right)}^2}}}$ is selected as the excitation, and uniform mesh size $ \Delta  = 5 \times {10^{ - 2}}$ m, and the simulation time is $t = 3.6685 \times {10^{ - 8}}$ s. Note that reflected electromagnetic waves do not reach probe in the simulations. 
\subsubsection{The FDTD(2,2) Method}
as discussed in the previous two sections, the numerical dispersion of the FDTD(2,2) method reaches its minimum value when $S_{fdtd(2,2)} = 1$ and it is called the magic time step \cite{TAFLOVE} in one dimensional case. The phase velocity $\tilde{v}_p$ exactly equals to $c_0$ in the free space. Therefore, the numerical results obtained from the FDTD(2,2) method do not suffer from the NDE. The magic time-step has already been proved by a rigorous mathematical manner in \cite{TAFLOVE}. Interested readers are referred to it for more details.

As shown in Fig. 7, when $S_{fdtd(2,2)} = 1$, the waveform obtained from the FDTD(2,2) method exactly agrees with the analytical solution. However, the other two results with $S_{fdtd(2,2)} = 0.7$ and $0.5$ show large discrepancy even though much smaller time steps are used in our simulations. In addition, it can be found that the error obtained from the FDTD(2,2) method when $S_{fdtd(2,2)} = 0.5$ is much larger than that with $S_{fdtd(2,2)} = 0.7$. It matches our analytical analysis in the previous two sections. 

\begin{figure}\label{one_fdtdtwo}	
	\centerline{\includegraphics[width=3.5in]{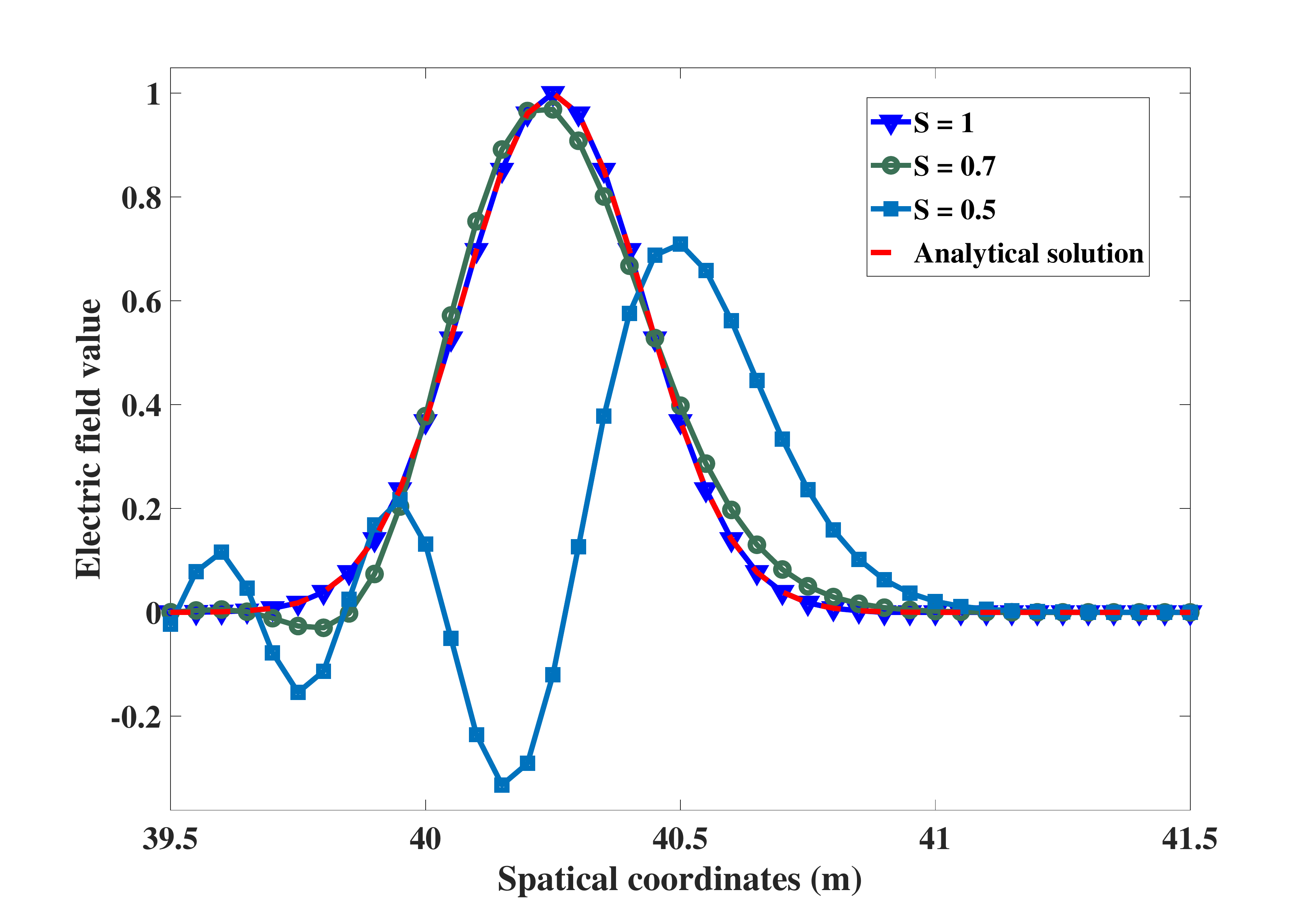}}
	
	\caption{Waveform obtained from the FDTD(2,2) method with different $S_{fdtd(2,2)}$.}
\end{figure}

\subsubsection{The FDTD(2,4) Method}
\begin{figure}\label{high_numerical}	
	\centerline{\includegraphics[width=3.5in]{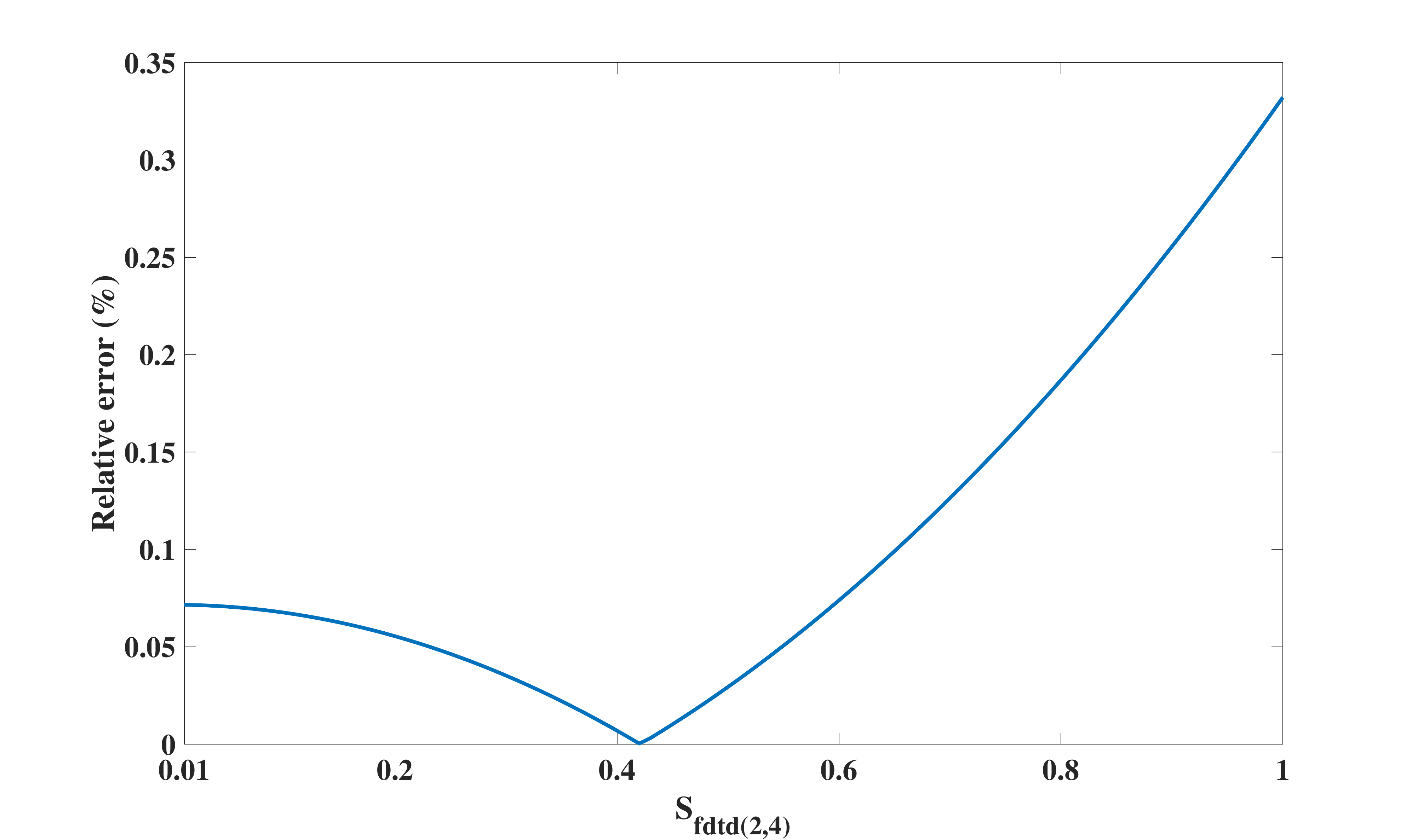}}
	
	\caption{Relative error of numerical dispersion of the one-dimensional FDTD(2,4) method with different $S_{fdtd(2,4)}$.	}
\end{figure}

Fig. 8 illustrates the absolute relative error with respect to $S_{fdtd(2,4)}$. It is easy to find that the relative error decreases as $S_{fdtd(2,4)}$ becomes larger from zero and reaches zero when $S=0.44$. Then, the relative error increases as $S_{fdtd(2,4)}$ continues to get larger. Since $\tilde{k}$ can be smaller or larger than $k$, the relative error shows different behaviors compared with that of the FDTD(2,2) method.  

Fig. 9 shows the waveform obtained from the FDTD(2,4) method. When $S_{fdtd(2,4)}=0.44$, the numerical result perfectly matches the analytical solution. However, other results obviously deviate from the analytical solution no matter when $S_{fdtd(2,4)}=1$ or $S_{fdtd(2,4)}=0.1$.

\begin{figure}\label{high_ed}	
	\centerline{\includegraphics[width=3.5in]{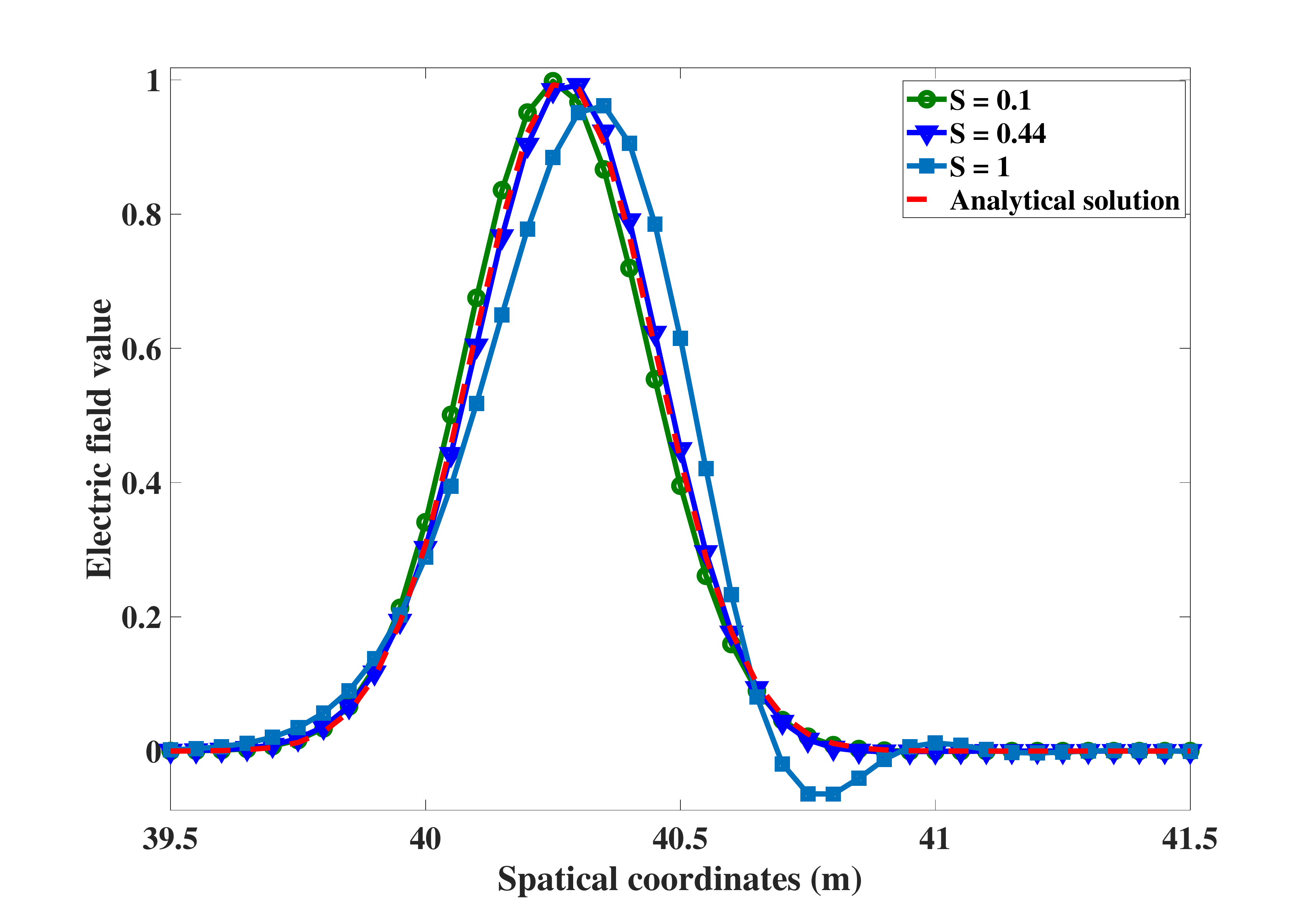}}
	
	\caption{Waveform of a Gaussian pulse obtained with the FDTD(2,4) method with different $S_{fdtd(2,4)}$ compared with the analytical solution.
	}
\end{figure}

\subsection{The Two Dimensional Case}
A two dimensional cavity with dimension of $1$ m $\times$ $2$ m and perfectly electric conductor (PEC) boundaries is considered and its resonant frequencies in both transverse magnetic (TM) and transverse electric (TE) modes are calculated through the two FDTD methods. The mesh size is $\Delta = 4 \times {10^{ - 2}}$ m. 
\subsubsection{The FDTD(2,2) Method}
 Fig. 10 shows the relative error of resonant frequencies of TM modes obtained from the FDTD(2,2) method. It's clear that despite of some numerical fluctuations, as $S_{fdtd(2,2)}$ increases, the relative error becomes smaller for three modes, which agrees well with our investigations. 

\begin{figure}\label{fdtd_Tm}	
	\centerline{\includegraphics[width=3.5in]{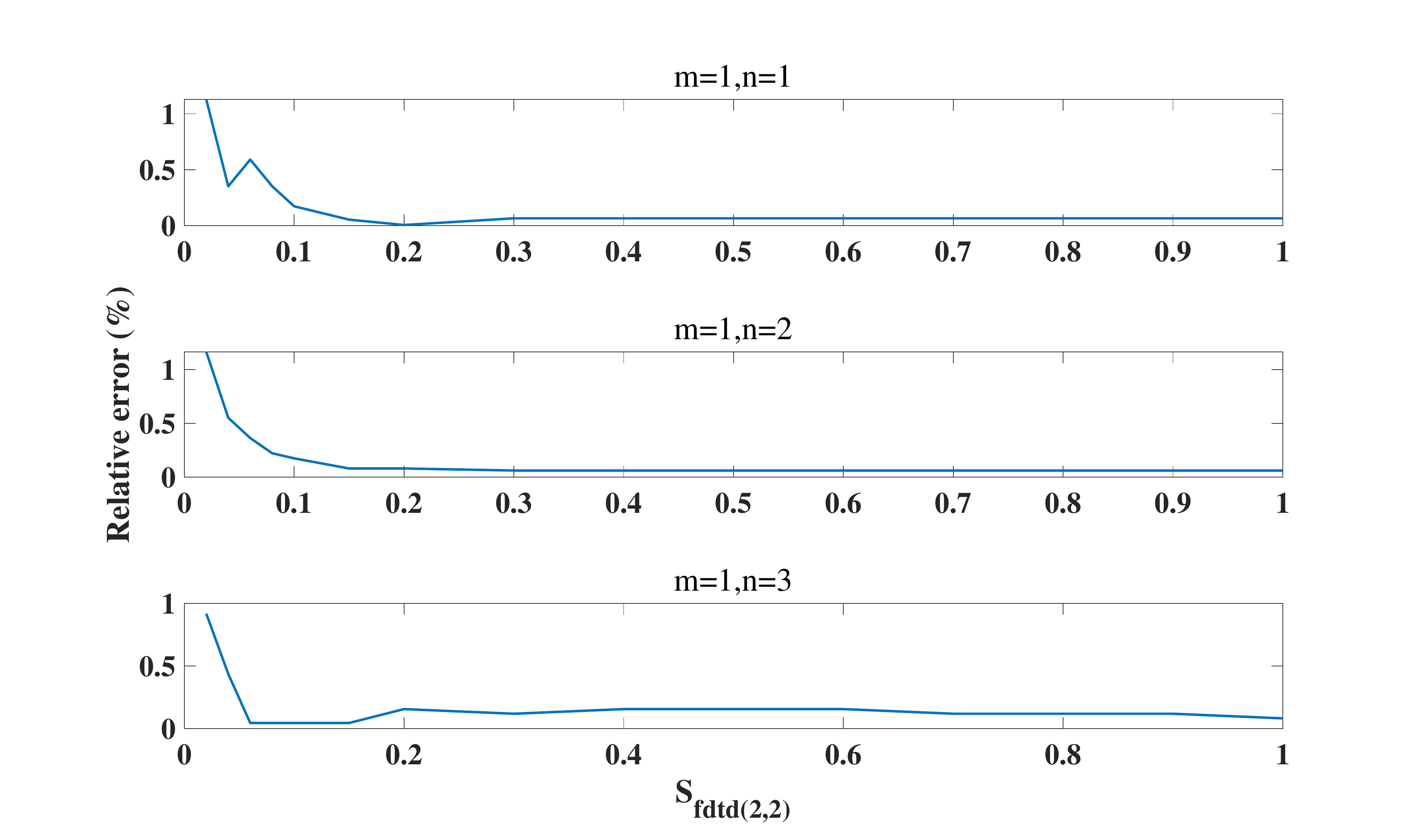}}
	
	\caption{Relative error of resonant frequencies for TM modes obtained from the FDTD(2,2) method with different $S_{fdtd(2,2)}$, $ m,n $ denote model number.
	}
\end{figure}

\begin{figure}\label{fdtd_two_Te}	
	\centerline{\includegraphics[width=3.5in]{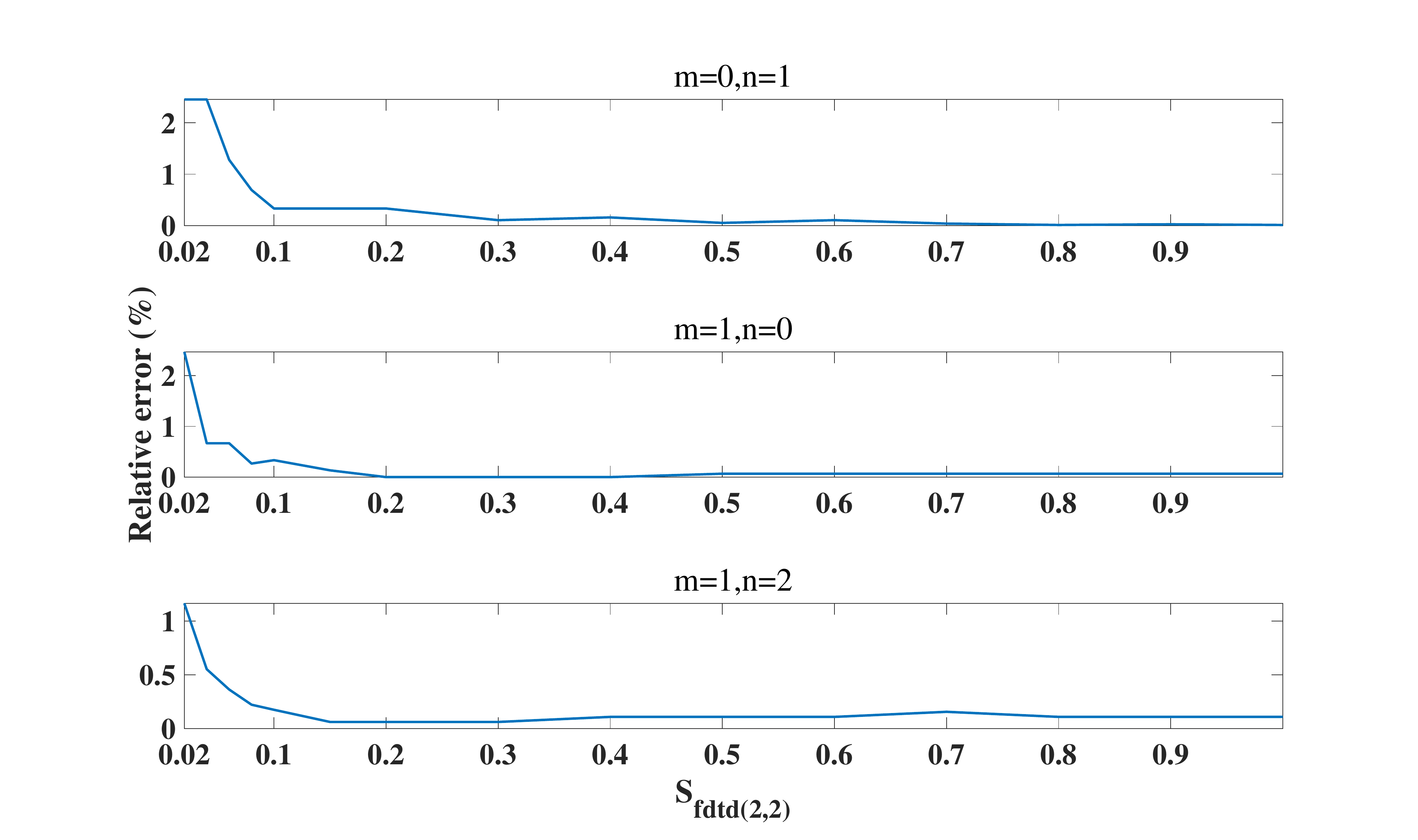}}
	
		\caption{Relative error of resonant frequencies for TE modes obtained from the FDTD(2,2) method with different $S_{fdtd(2,2)}$, $ m,n $ denote model number.
	}
\end{figure}

Fig. 11 shows relative errors of TE modes. Similar observations can be made as to TM modes. When $S_{fdtd(2,2)}$ grows larger, the NDE in the FDTD(2,2) method decreases gradually and more accurate simulation results can be obtained. Therefore, the optimum time step is $S_{fdtd(2,2)}=1$.

\subsubsection{The FDTD(2,4) Method}
the resonant frequencies of TM and TE modes obtained from the FDTD(2,4) method are shown in Fig. 12 and Fig. 13.  It is easy to find that the relative errors of both three TE and TM modes show similar behaviors. The relative error gradually decreases and then slightly becomes larger as $S_{fdtd(2,4)}$ become larger. It should be noted that the relative error with small $S_{fdtd(2,4)}$ is larger than that with large $S_{fdtd(2,4)}$ as shown in Fig. 12 and Fig. 13. It may account for other numerical errors, like low order approximation of boundaries. Note that compared with the results in Fig. 10 and Fig. 11, results obtained from the FDTD(2,4) method are obviously more accurate than those obtained from the FDTD(2,2) method.

\begin{figure}\label{TM_high}	
	\centerline{\includegraphics[width=3.5in]{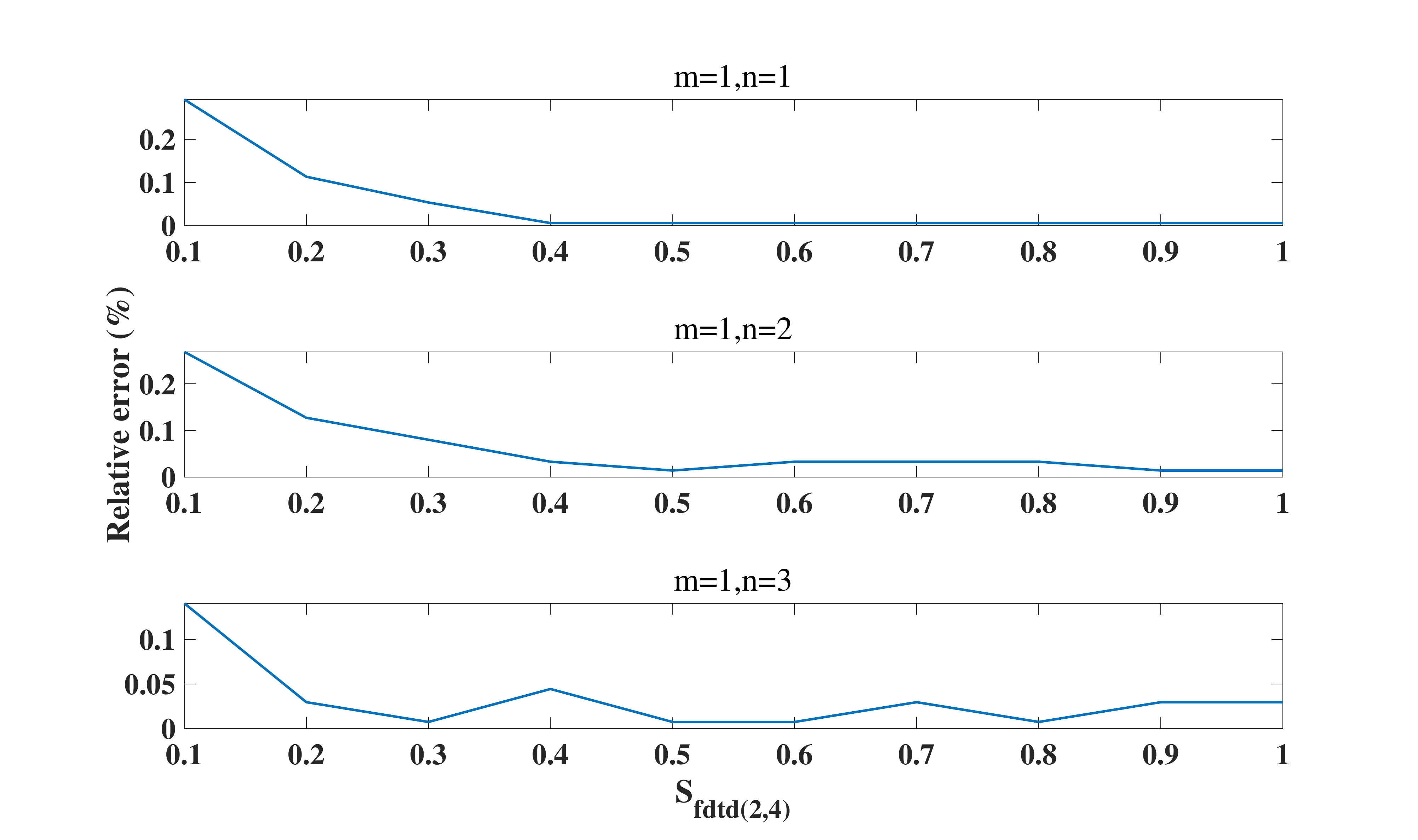}}
	
	\caption{Relative error of resonant frequencies for TM modes obtained from the FDTD(2,4) method with different $S_{fdtd(2,4)}$, $ m,n $ denote model number.
	}
\end{figure}

\begin{figure}\label{TE_high}	
	\centerline{\includegraphics[width=3.5in]{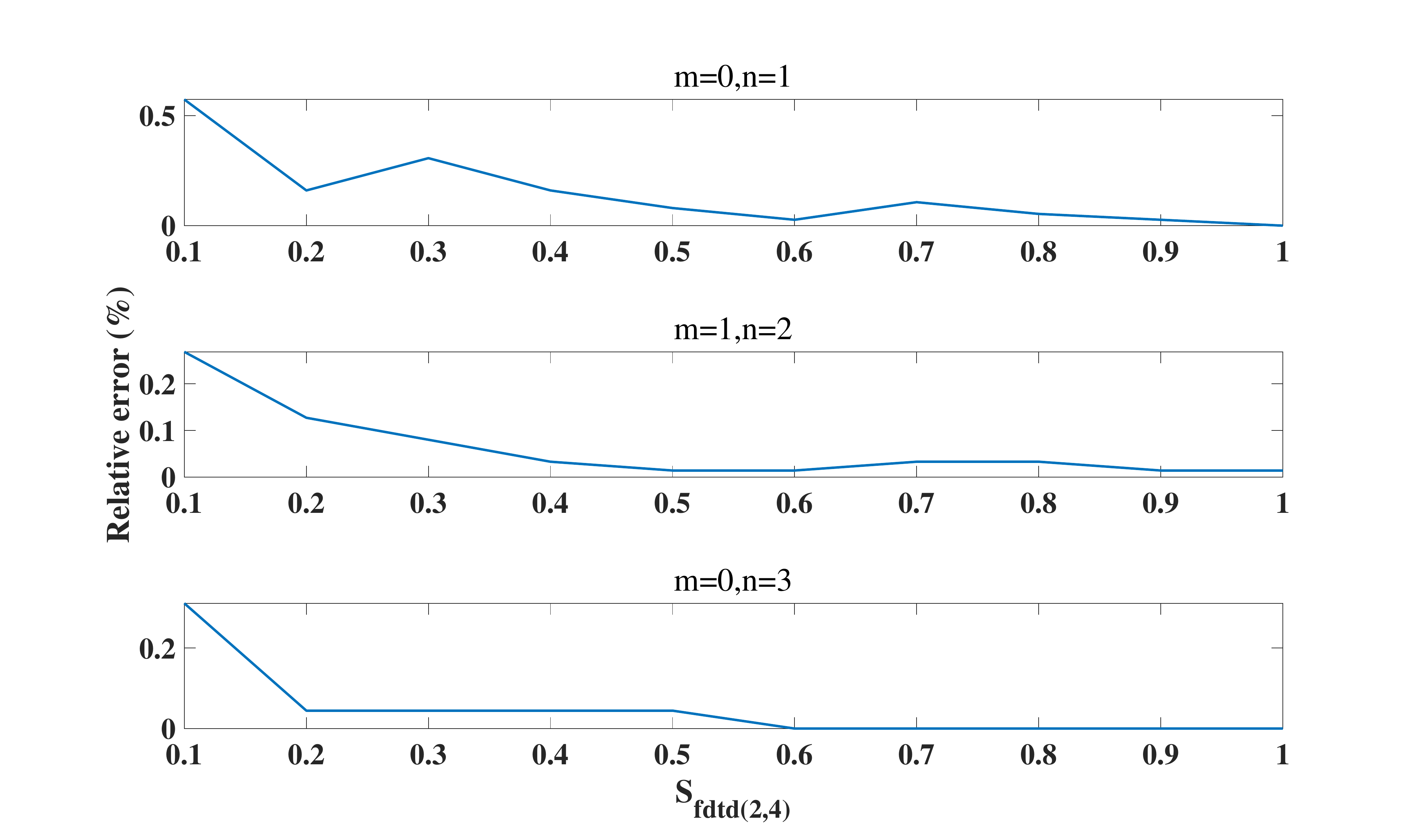}}
	
	\caption{Relative error of resonant frequencies for TE modes obtained from the FDTD(2,4) method with different $S_{fdtd(2,4)}$, $ m,n $ denote model number.
	}
\end{figure}

\subsection{The Three Dimensional Case}
{\bf{Case A:}} a cubic cavity with side length of $1$ m in the three dimensional case is considered. The mesh size is $\Delta = 4 \times {10^{ - 2}}$ m. The resonant frequencies are calculated from the two FDTD methods.

\begin{figure}\label{fdtd_three}	
	\centerline{\includegraphics[width=3.5in]{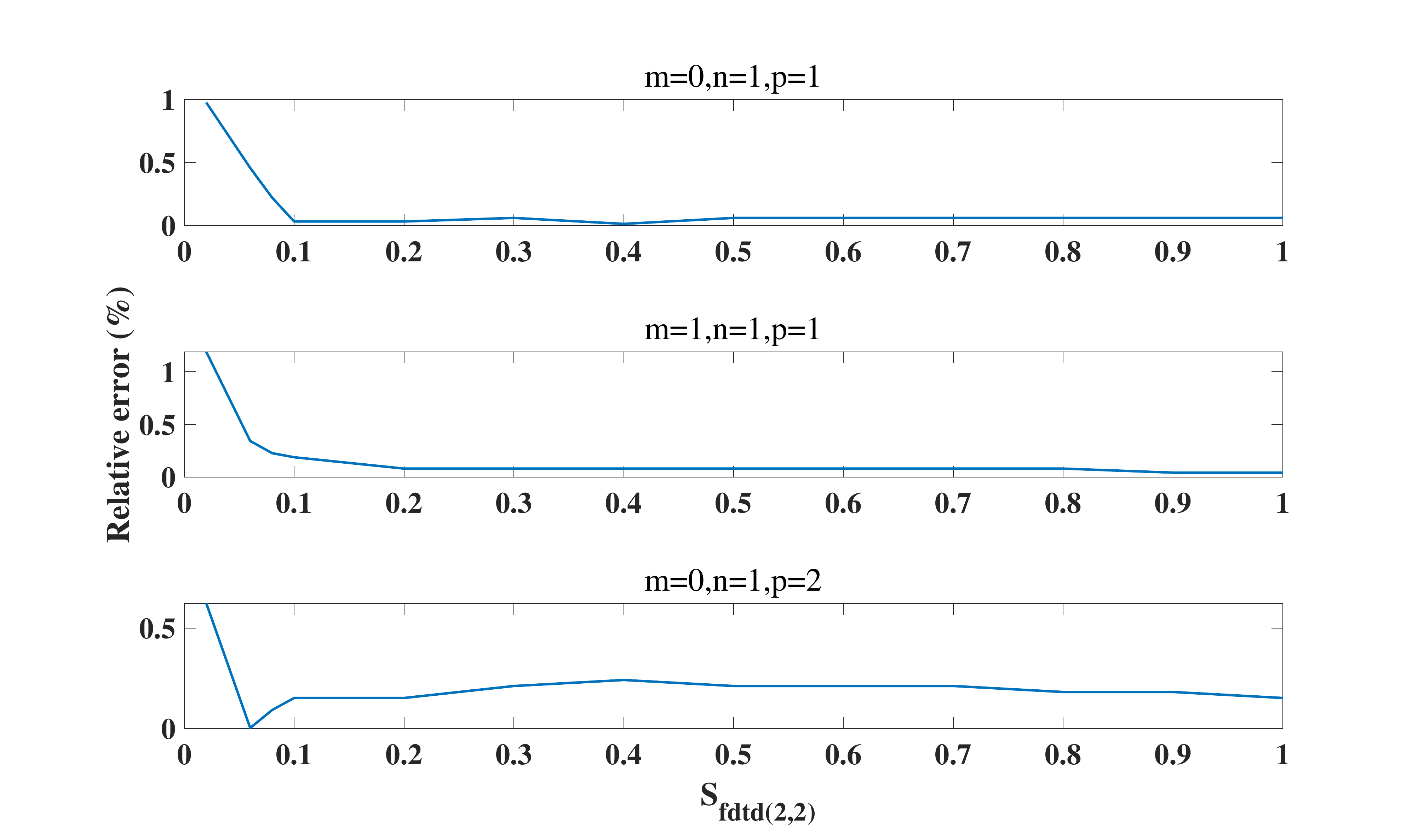}}
	
	\caption{Relative error of resonant frequencies obtained from the FDTD(2,2) method with different $S_{fdtd(2,2)}$, $ m, n, p$ denote model number.
	}
\end{figure}

Fig. 14 shows the relative error of resonant frequencies obtained from the FDTD(2,2) method. It is similar to one- and two-dimensional cases. Although there are some numerical fluctuations, the relative error becomes smaller for all the three modes as $S_{fdtd(2,2)}$ increases. It agrees well with our investigations.

\begin{figure}\label{three_high}	
	\centerline{\includegraphics[width=3.5in]{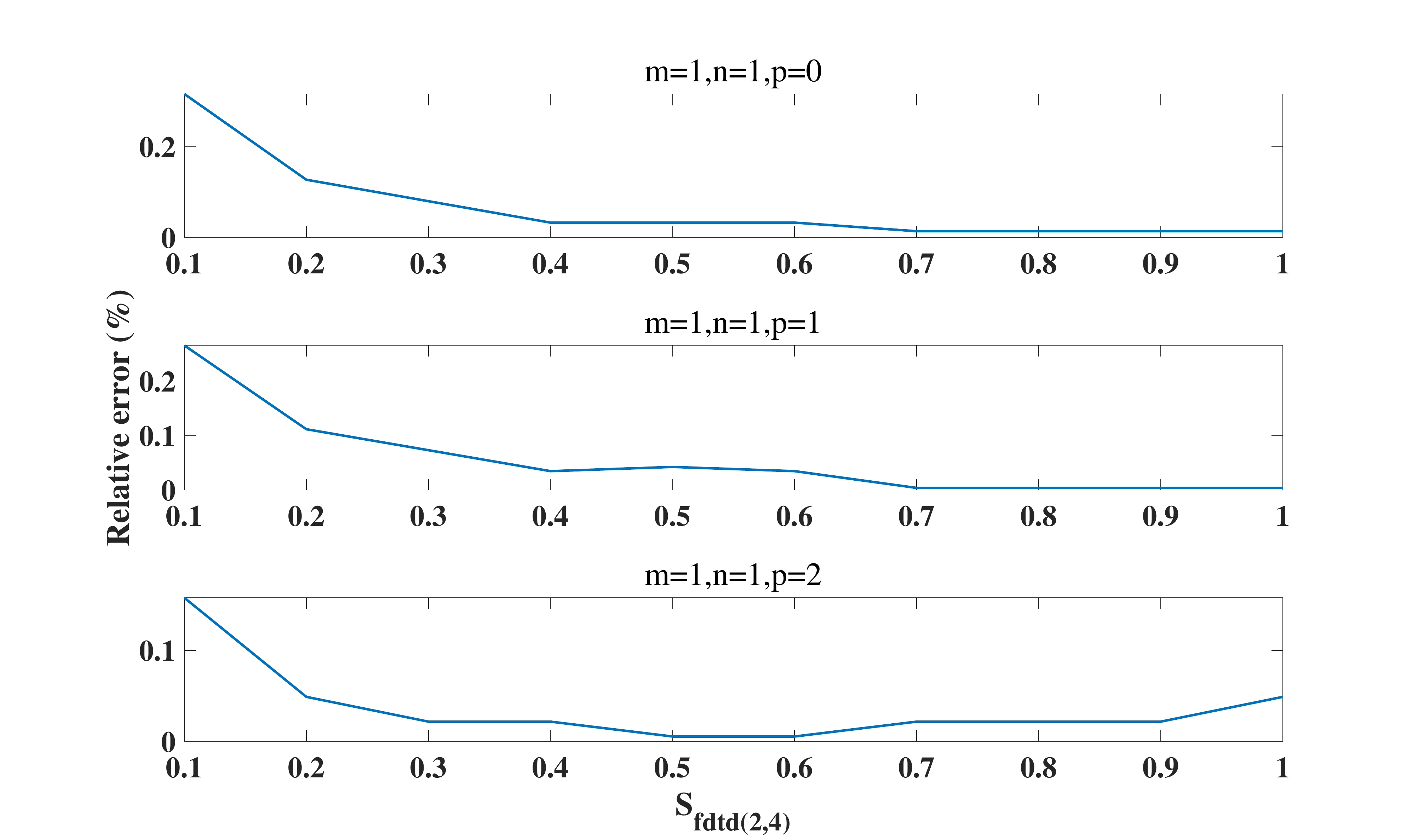}}
	
	\caption{Relative error of resonant frequencies obtained from the FDTD(2,4) method with different $S_{fdtd(2,4)}$, $ m, n, p$ denote model number.
	}
\end{figure}

Fig. 15 shows the resonant frequencies obtained from the FDTD(2,4) method.  It is easy to find that relative errors of resonant frequencies show similar behaviors. The relative error gradually decreases and then slightly becomes larger as $S_{fdtd(2,4)}$ become larger. The relative error with small $S_{fdtd(2,4)}$ is larger than that with large $S_{fdtd(2,4)}$. It may account for other numerical errors.

\begin{figure}
	\begin{minipage}[h]{0.48\linewidth}\label{coppermodel+RCS}
		\centerline{\includegraphics[scale=0.2]{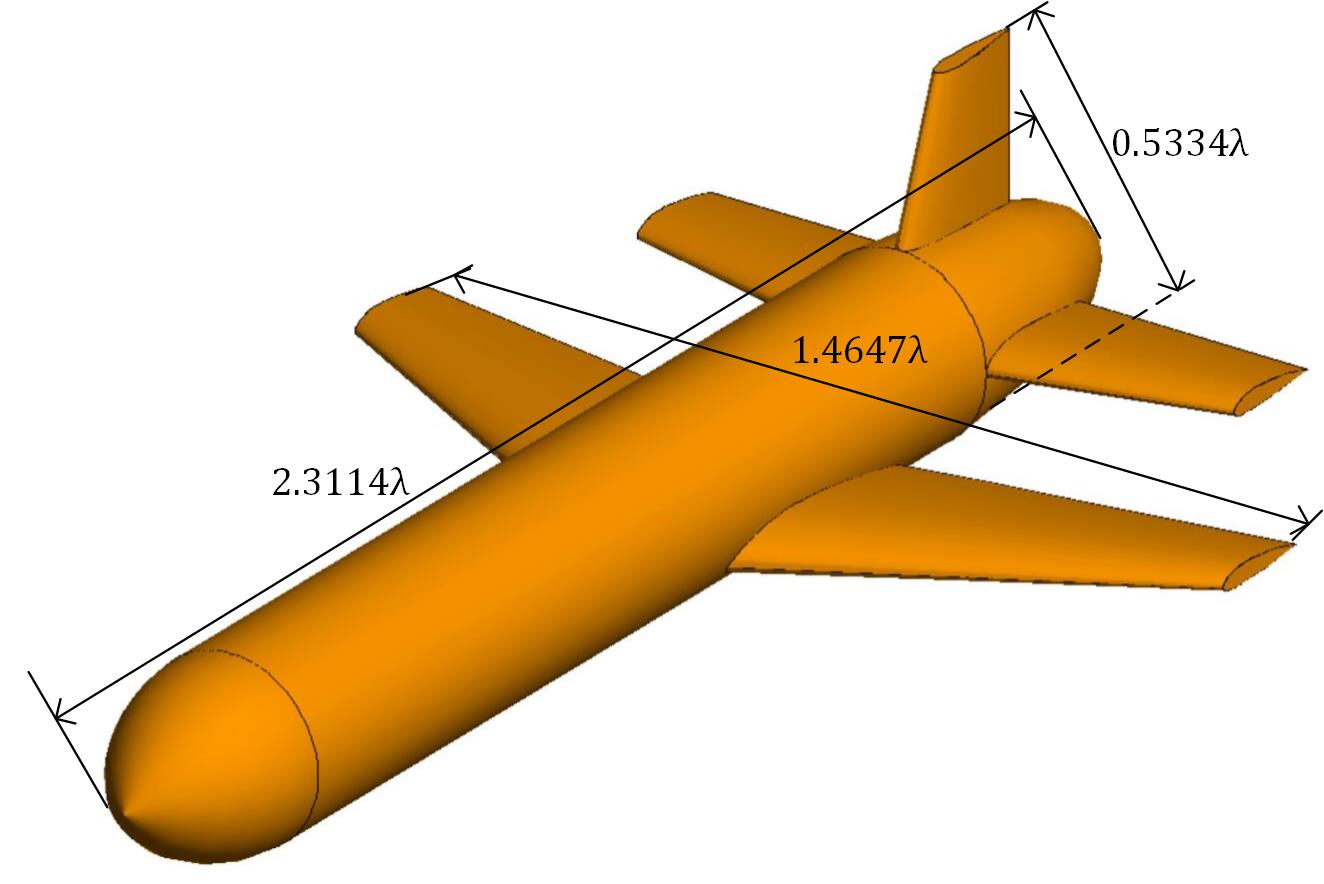}}
		\centerline{(a)}
	\end{minipage}
	\centering
	\vfill
	\begin{minipage}[h]{0.48\linewidth}\label{copperRMS}
			\centerline{\includegraphics[scale=0.09]{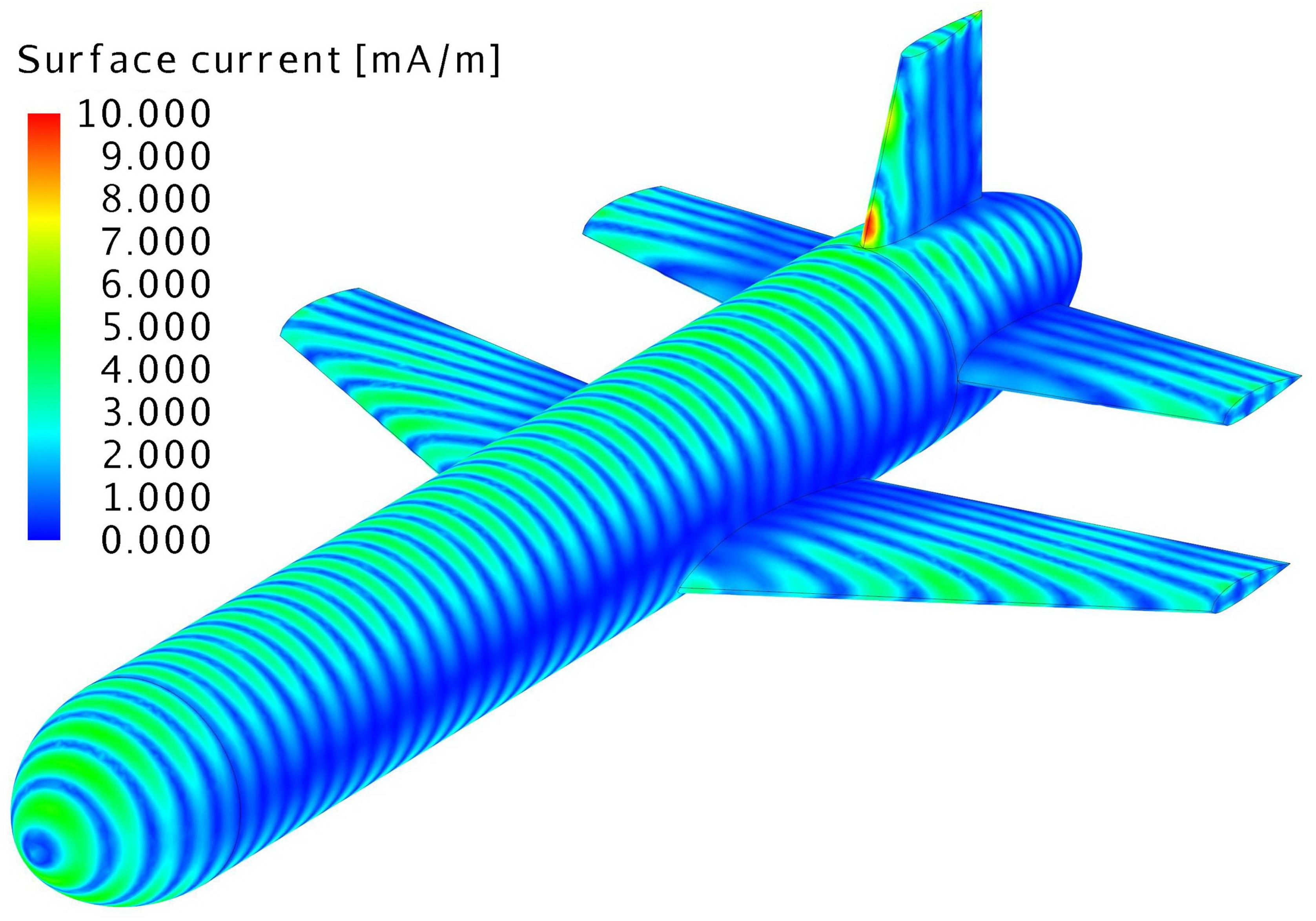}}
		\centerline{(b)}
	\end{minipage}
	\centering
	\caption{(a) Geometry configures of the missile illuminated from a plane wave with $f$=700 MHz, and (b) surface current density obtained from the FEKO.}
	\label{coppermodel+RCS}
\end{figure}

\begin{figure}
	\begin{minipage}[h]{0.48\linewidth}\label{coppermodel+RCS}
		\centerline{\includegraphics[scale=0.08]{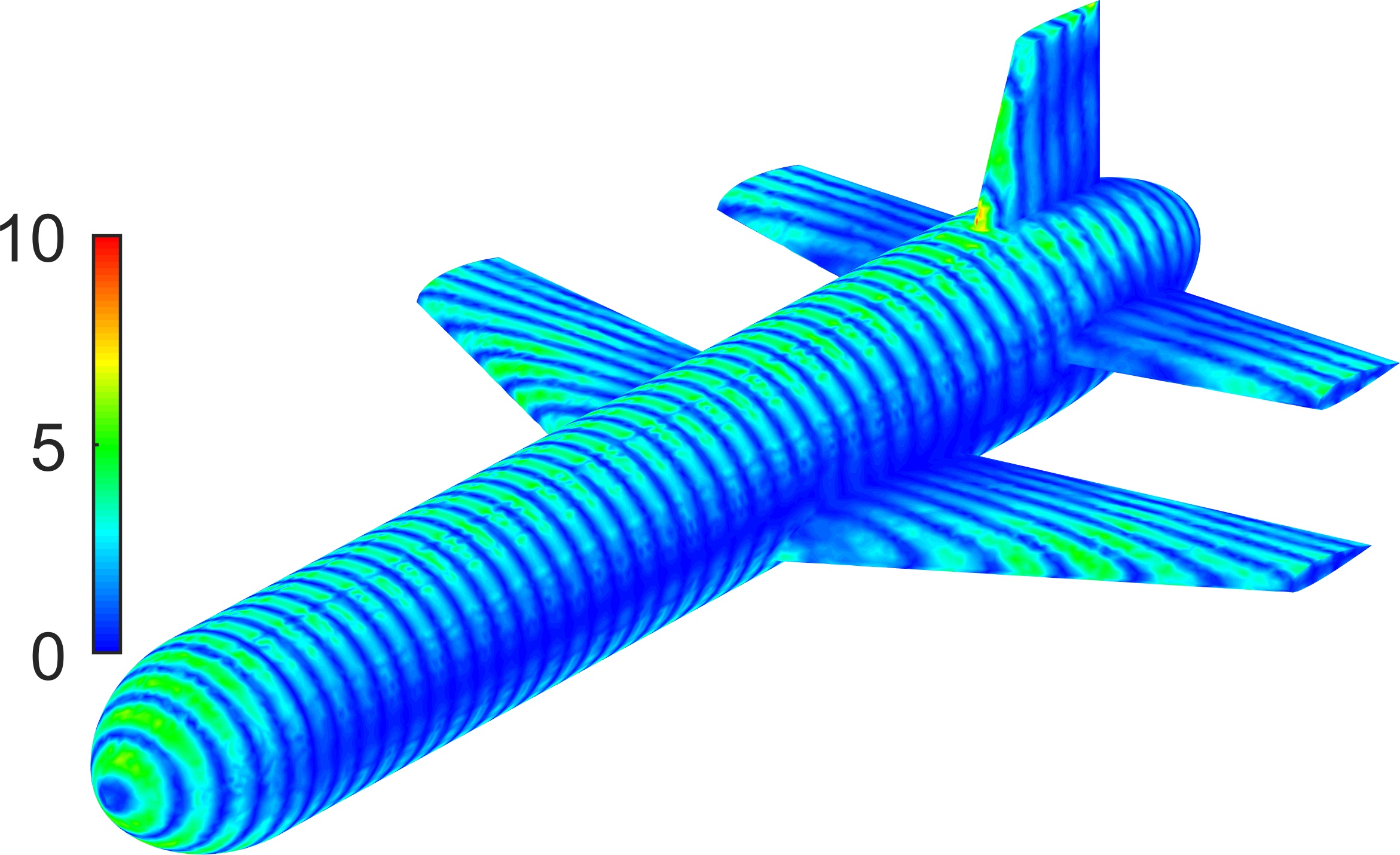}}	
		\centerline{(a)}
	
	\end{minipage}
	\centering
	\vfill
	\begin{minipage}[h]{0.48\linewidth}\label{copperRMS}
		\centerline{\includegraphics[scale=0.08]{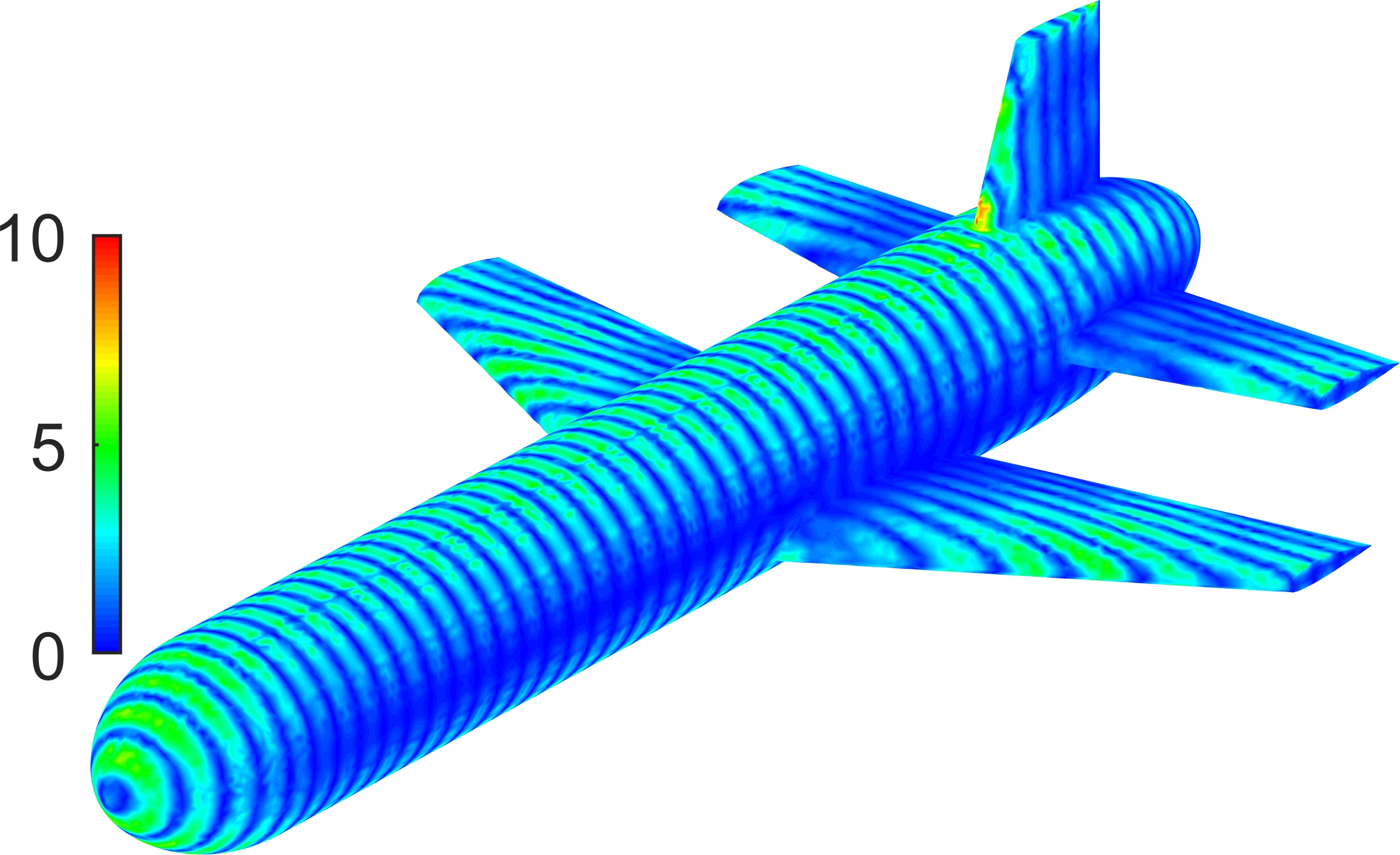}}	
		\centerline{(b)}
	
	\end{minipage}
     \centering
	\caption{(a) Surface current density obtained from the FDTD(2,2) method  with $S_{fdtd(2,2)} = 0.1$, and (b) $S_{fdtd(2,2)} = 1$.}
	\label{coppermodel+RCS}
\end{figure}

{\bf{Case B:}} we use the FDTD(2,2) method to further verify our findings for a practical application in this case. A missile is illustrated by a plane wave and the surface current density is calculated from the FDTD(2,2) method with different time steps. The geometry configurations of the missile in our simulations are shown in Fig. 16(a). It is with dimension of 2.3114$\lambda$$\times$1.4647$\lambda$$\times$0.5334$\lambda$. The plane wave incidents from the $x$-axis with 700 MHz. Fig. 16(b) shows the reference surface current density calculated from the FEKO.  Fig. 17(a) and (b) show the surface current density obtained in the FDTD(2,2) method with $S_{fdtd(2,2)} = 0.1$ and $1$. It can be found that the surface current distribution obtained from the FDTD(2,2) method agree well with the reference solution. The accuracy of results in our simulations seem unchanged with $S_{fdtd(2,2)} = 0.1$ and $1$. There are various factors that can account for those results in this case, like complex geometry, staircase error, possible reflection from the total field/scattered field boundary. It turns out that time steps of the FDTD(2,2) method almost have no significant effects on the accuracy in the practical simulation. Therefore, large time step is preferred in the practical  simulations since short simulation time is required and almost the same level of accuracy can also be achieved.

\section{Conclusion}
In this paper, we comprehensively investigated how time steps affect the accuracy of the FDTD methods in terms of numerical dispersion. Several findings are reported in this paper. Our results show that for the FDTD(2,2) method, smaller time step limited by the CFL condition leads to larger NDE. However, for the FDTD(2,4) method, as time step increase, the NDE first decreases and then increases. However, large time step of the FDTD method is preferred in
the practical simulations as shown in our numerical results, which means shorter simulation time.

The findings in this paper not only can further deepen our insights upon the FDTD methods and provide the guidance for selection of optimal time steps in the FDTD simulations, but also correct widespread erroneously thought about effects of time steps on numerical dispersion of different FDTD methods.

\end{document}